\documentclass[12pt,preprint]{aastex}

\slugcomment{}

\shorttitle{Co-evolution of AGN and their host galaxies}
\shortauthors{Reviglio, Helfand}

\begin{document}

\title{Active Galaxies in the Sloan Digital Sky Survey II: galaxy and activity evolution}

\author{Pietro M. Reviglio,  David J. Helfand}
\affil{Astronomy Department, Columbia University, New York, NY 10027}

\email{reviglio@astro.columbia.edu,djh@astro.columbia.edu}

\begin{abstract}
In this second paper of a series of papers based on the FIRST and SDSS surveys we investigate the evolution of galaxy morphology and nuclear activity  in the look-back time of the SDSS ($\sim$ 2 Gyrs) for a sample of ~150000 galaxies in the local universe.
We  demonstrate an evolution in the strength of the radio power and the spectroscopic emission-lines typical of AGN, as well as in the morphology of their hosts. Such evolution appears more substantial for less luminous systems, and is possibly the low-redshift tail of the downsizing in  star-formation, AGN activity and supermassive black hole build-up observed in higher redshift surveys.This suggests that the differences in  intrinsic properties  of galaxies along the Hubble Sequence  may arise from the difference in the depth of their potential wells which leads to  different evolutionary paths because of different timescales for  gas infall. 
This primordial infall and the following secular evolution mediated by bar and density wave instabilities may bring galaxies of different mass to have the different activity levels and morphological features in the local universe shown in this study.   In agreement with such a hypothesis, we find that star-formation as traced by radio emission is progressively more centrally concentrated in more evolved star-forming galaxies and we show that the environment in which a  galaxy resides plays a lesser role in shaping the features and activity for the majority of systems.

\end{abstract}


\section{Introduction}
\label{sec:intro}

Understanding the activity of galaxies, their morphologies, and the interplay between the two are key problems in galaxy evolution and are tightly correlated with the problem of understanding whether such properties depend on the initial physical conditions of a galaxy (e.g., its mass) or, rather,  result from evolutionary processes driven by the environment in which they live, such as mergers, gravitational and gas-gas  interactions.

Models of galaxy formation and evolution have tried to explain the differences in the features of galaxies  for decades, mostly unsuccessfully. 
Hierarchical cold dark matter models, which have been shown to describe successfully large-scale structure formation, currently fail to describe properly  many salient properties of galaxies, such as the color-magnitude relation, the bimodal distribution in colors, the luminosity function, and ``downsizing" (i.e., the earlier quenching of star formation in  more massive objects). Furthermore such models  inadequately describe the formation of small bulge-less spirals, the most common galaxies in the universe. It is unclear if the problems that these models encounter are given by a simplistic  treatment of the complex baryonic physics governing the observable parameters or if the models are not appropriate for the  description of galaxy formation and evolution.

 Explaining the mechanisms which initiate and regulate star-formation and AGN activity in galaxies along the Hubble diagram has proven as challenging. Galaxies show very different rates and histories of star-formation activity and host AGN with very different characteristics (narrow and/or broad emission lines, point-like radio emission and large radio jets, to name a few), making the modeling of such activity quite complex. 

The interplay between star-formation and AGN activity is also poorly understood. Using the SDSS, \cite{Kauffmann2003} showed that active galactic nuclei selected on the basis of their emission lines preferentially reside in massive,  bulge-dominated systems with traces of recent star-formation, possibly a consequence of the presence of a reservoir of cold gas  which triggers both star-formation and AGN activity.  Using the GOODS Survey, however, \cite{Nandra2007}  showed  that X-ray AGN activity persists in many $z \sim 1$ red-sequence galaxies, well after the star-formation has quenched, suggesting that the presence of a cold reservoir of gas may not be a necessary ingredient in the  triggering of AGN activity.

 The role of the environment surrounding a galaxy in initiating activity is also quite controversial.
 Interactions and mergers among galaxies are known to trigger bursts of star formation in galaxies \cite[e.g.,][]{Barton2000}, yet the trigger for the onset of star-formation on galactic scales in isolated spiral galaxies calls for a different explanation.
In clusters, \cite{Christlein2005} have shown that the environment produces a significant effect in the later evolution of galaxies, lowering their star-formation rates substantially,  while  \cite{Hogg2006} using the SDSS sample conclude the opposite, i.e. that the cluster environment may be irrelevant to the bulk of the transformation of galaxies. 
Mergers and close interaction between galaxies have also been suggested as possible triggers of AGN activity \cite[]{Dahari1984}; however,  \cite{Grogin2005} using the GOODS survey and two Chandra Deep Fields have shown that there is no close connection between galaxy mergers and moderate-luminosity AGN activity out to appreciable look-back times approaching the epoch of peak AGN activity in the universe ($z\le 1.3$). This result is supported by the study of \cite{Pierce2007} who showed that X-ray-selected AGN tend to reside in undisturbed pairs of early type galaxies with no direct evidence of gravitational perturbation or interactions, suggesting that the activation mechanism of AGN is not primarily galaxy interaction.

The role of secular evolution in triggering star-formation and AGN activity, as well as the morphological transformation of galaxies by means of a re-arrangement of energy and mass through collective phenomena such as bars \cite[]{Combes2000} and/or spiral waves \cite[]{Zhang1996,Sellwood2002} is unclear, but there is increasing observational evidence of its importance in shaping galaxy features  \cite[]{Kormendy2004}.

Interestingly, many properties observed in the local universe are observed also in the high- redshift population of galaxies investigated by recent deep galaxy surveys. Using the GOODS survey, \cite{Grogin2005} have shown that the well-known correlation between size of a galactic bulge and the mass of the supermassive black hole residing at its  center \cite[]{Ferrarese2000} is in place at early epochs (z$\le$1.3).   \cite{Cassata2007}, using the COSMOS Survey, have studied the morphological content and environmental dependence of the galaxy color-magnitude relation at $z \sim 0.7$ and found that the red and blue sequences of galaxies found in the nearby universe are found at $z \sim 0.7$ as well. By comparing the high- and low-redshift populations, they conclude that the local red sequence is adequately reproduced by passive evolution, suggesting that a fraction of the galaxies in the high-redshift blue sequence  eventually evolve into red-sequence galaxies. They also confirm the result of \cite{Balogh2004}, who showed that the slope and the normalization of such color-magnitude relations are substantially invariant with respect to local density when galaxies are separated into early and late types, arguing that secular star-formation is driven more by galaxy mass than by environment. The color-density correlation found by other authors \cite[e.g.,][]{Cooper2008} at high redshift, with bluer galaxies found in regions of greater over-density, is interpreted as an excess of more late-type galaxies in denser environments which evolve into red-sequence galaxies at the present time. These studies are in  agreement with older results \cite[]{Butcher1978}  which showed a variation in the fractional abundance of different morphological types in galaxy  cluster populations as a function of  redshift. The study of \cite{Cassata2007}   is not limited to clusters, suggesting that a transformation of galaxies might  also occur  in the absence of  mechanisms such as the gas-gas and gravitational interactions typical of dense cluster environments.
These results are at odds with the work of \cite{Park2007} who have used the SDSS sample to show that close interaction among galaxies is the most likely mechanism leading to the density-morphology correlation observed.

In a series of three papers  based on the Sloan Digital Sky Survey \cite[]{York2000} and the FIRST Survey \cite[]{Becker95} we examine the population of broad and narrow emission-line active galactic nuclei (AGN) in the local universe in order to explore the  physical and environmental properties of galaxies.

 In Paper I \cite{Reviglio2009} we described the multiwavelength database built for this analysis and the systematics affecting the spectroscopic selection of AGN. We presented evidence of the existence of a large population of radio-emitting radio AGN which lack optical spectroscopic signatures and show that, at all luminosities, galaxies with redder colors are more likely to show weaker emission lines. In this second paper we further explore the interplay between galaxy properties and AGN features, focusing on the interplay between morphology, star-formation, and nuclear activity, including its evolution over cosmic times. In a following paper we will present a comparative analysis of the radio and spectroscopic properties of narrow- and broad-lined AGN and their evolution.

The structure of the paper is the following:  in section \ref{sample_ch3} we present the sample used for this analysis, while in \S  \ref{evolution} we show our evidence of co-evolution of  galaxy activity and morphology, and in \S \ref{aging_radio} we  map the aging of the radio power associated with AGN. In \S \ref{aging_spectro} we illustrate the aging of the spectroscopic features of AGN, in \S \ref{environment} we study the correlation between environment, activity and morphology. We discuss our results in \S \ref{discussion_ch3} and present our conclusions in \S \ref{conclusion_ch3}.

\section{The sample}
\label{sample_ch3}
In this paper we further analyze a sample of 151815 galaxies with Petrosian R magnitude in the range 14.5$\le$R$\le$17.5 drawn from the SDSS Second Data Release, for which we have constructed a multiwavelength database, gathering information about their radio emission from the FIRST and NVSS surveys, the  far-infrared emission from IRAS, and x-ray emission from the ROSAT surveys.
Details on the database construction are given in Paper I.
The density of the environment surrounding each galaxy has been  calculated using the tridimensional density estimator discussed in Paper I; further description of this method can be found in \cite{Carter01}. 

The spectroscopic classification of galaxies is adopted  from \cite{Kauffmann2003} and \cite{Brinchmann2004}, along with  the extinction-corrected fluxes for the main spectroscopic lines, their equivalent widths,  and the extinction-corrected estimates of the stellar masses of the galaxies in the sample. As discussed in Paper I, we have modified the classifications to correct for effects of dilution and host light pollution which tend to bias the classifications with increasing redshifts.

About 6\% of the sample shows detectable radio emission in either FIRST or NVSS.
For these radio-emitting galaxies, the spectroscopic classification has been complemented with an independent classification based on the radio and optical morphology of the sources, as discussed in Paper I and in \cite{Reviglio2008}.

\section{Evolution of AGN activity and host properties}
\label{evolution}

The look-back time of the SDSS is about 2 Gyrs in standard cosmological models ($\Omega_m=0.3$ and $\Omega_\Lambda=0.7$). The typical timescales for the processes associated with AGN activity and star-formation are much shorter, typically a few tens of millions of years. Therefore we wish to investigate whether evolution of activity and morphology can be detected over the  look-back time of the SDSS.

 In order to compare the properties of galaxies at different redshifts it is crucial to minimize selection effects given by the survey systematics.
In Paper I we showed how dilution significantly biases the selection of active galaxies with increasing redshift, and devised a method to correct for this bias. This correction is essential, since dilution is a redshift-dependent systematic and if not properly corrected, could mimic evolution of the sample with redshift (cf. Paper I for a more detailed discussion).

\begin{figure}[h]
\begin{center}
\includegraphics[scale=0.5, angle=90]{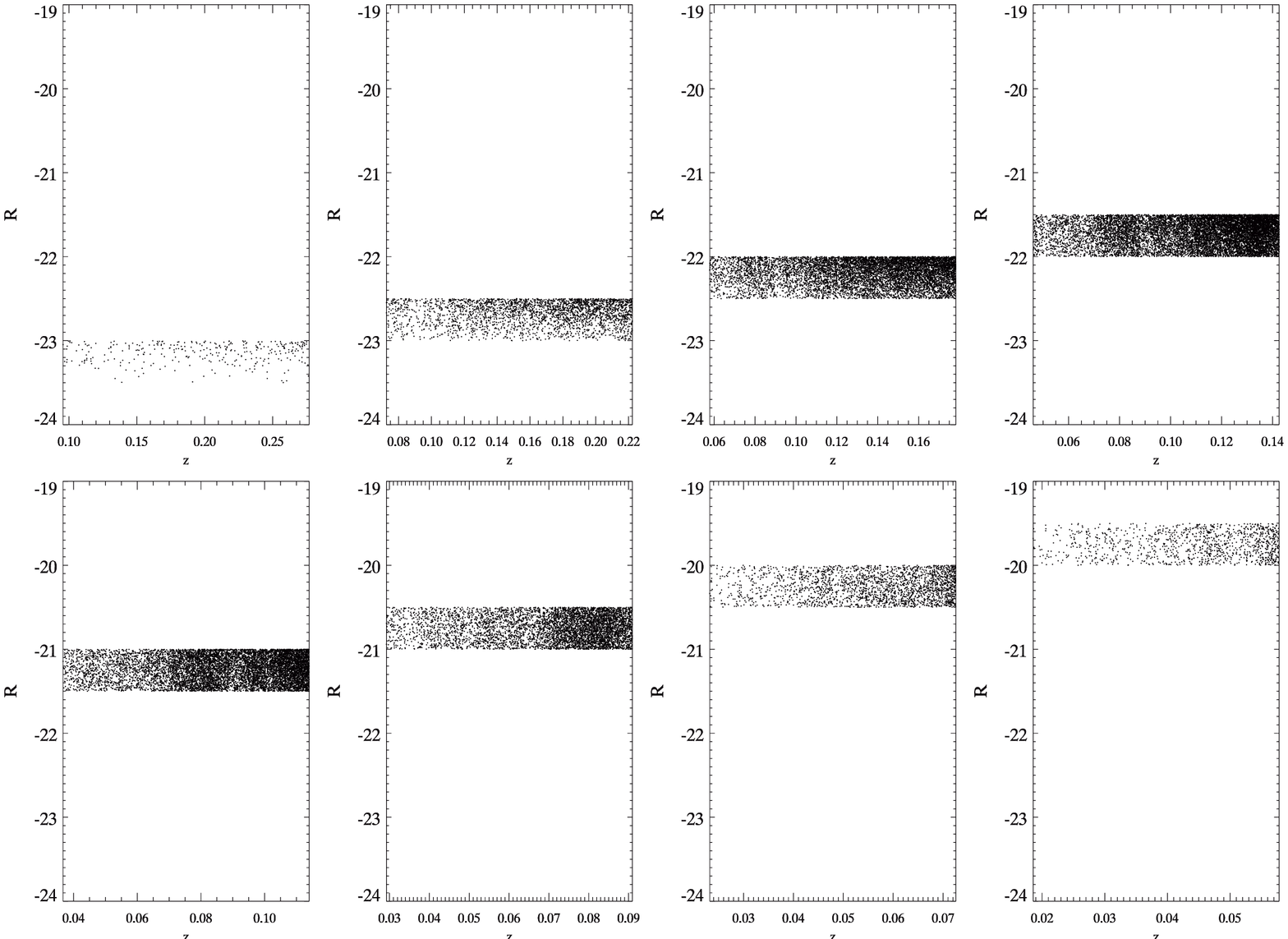}
\end{center}
\caption{\footnotesize \emph{The eight volume-limited samples used in this analysis.}}
\bigskip
\label{abs_r_AGN}
\end{figure}

In order to explore the properties of AGN and host galaxies over the look-back time  of the SDSS, we have focused on volume-limited samples of galaxies which ensure completeness of the optical sample within the redshift and absolute magnitude cuts applied. We have constructed a chain of eight volume-limited samples covering the absolute magnitude range $-23.5<R<-19.5$ at intervals of 0.5 in magnitude. This ensures that we are looking at optically complete samples of AGN hosts with similar properties. In figure \ref{abs_r_AGN}  we show the scatter plot in redshift and absolute magnitude of galaxies belonging to each volume-limited sample.

\begin{figure}[h]
\begin{center}
\includegraphics[scale=0.5, angle=90]{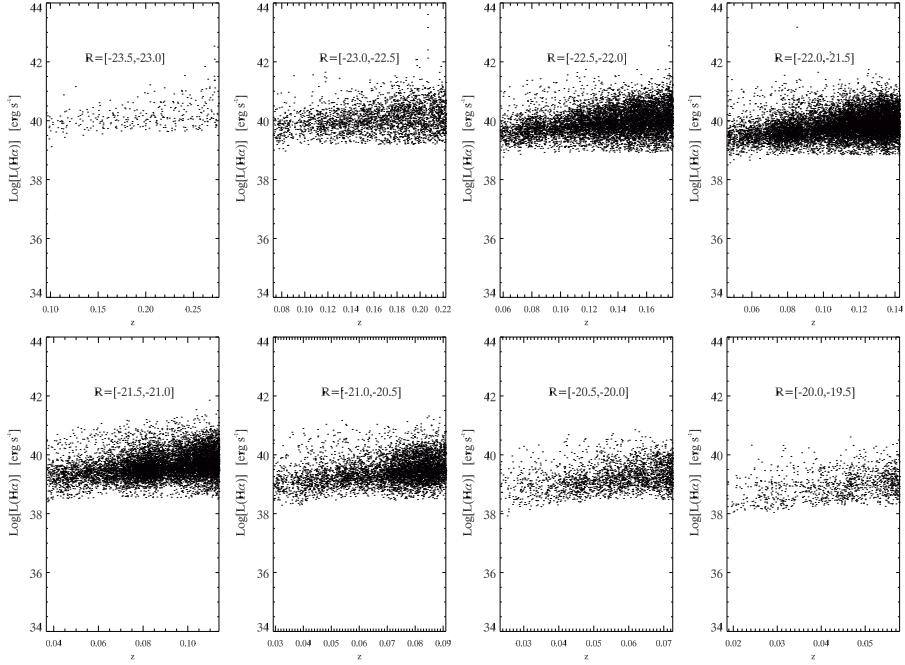}
\end{center}
\caption{\footnotesize \emph{The luminosity in H$\alpha$ of emission-line AGN in the eight volume-limited samples considered in this study. The sample has been corrected for dilution, as discussed in Paper I.}}
\bigskip
\label{Lum_ha_z_AGN}
\end{figure}

For the AGN-classified galaxies, we show in Figure \ref{Lum_ha_z_AGN} the scatter plot of  H$\alpha$ luminosity versus the luminosity distance, and we note how the H$\alpha$ luminosity decreases over time. The effects of dilution have been corrected for (no lower cut appears in the plots)  but even if they had not, this would not account for the lack of strong emission-line AGN at lower redshifts.

\begin{figure}[h]
\begin{center}
\includegraphics[scale=0.5, angle=90]{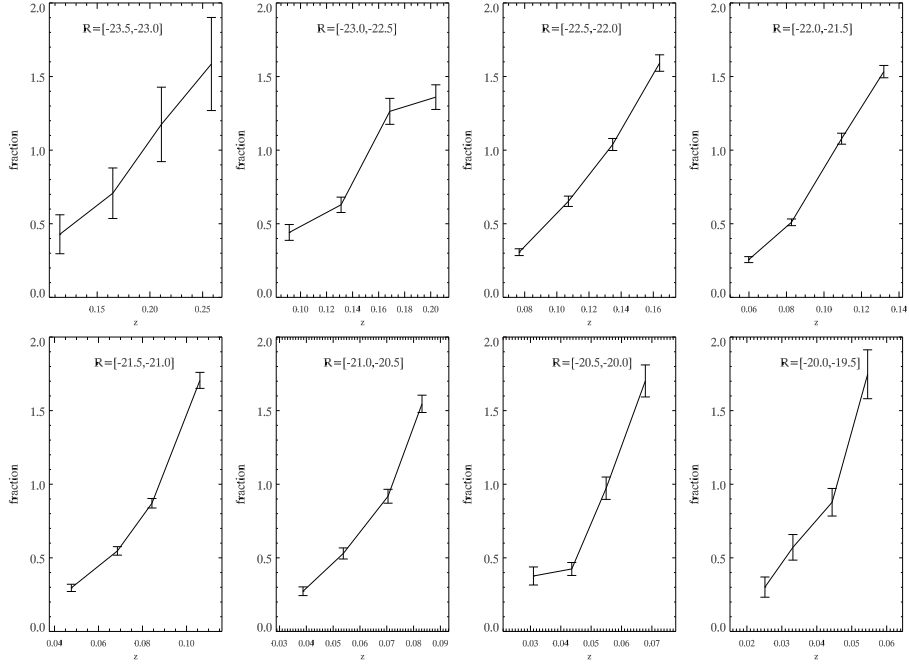}
\end{center}
\caption{\footnotesize \emph{The ratio of spectroscopic AGN with luminosity in H$\alpha$ above and below the median in the eight volume-limited samples considered in this study. AGN evolve toward lower H$\alpha$ luminosities}}
\bigskip
\label{Lum_ha_z_AGN_frac}
\end{figure}

Differences in the volumes sampled at various redshifts, however, might explain the lack of high-luminosity systems at lower redshifts. A better way to explore this trend is to consider for each volume-limited sample the ratio of galaxies with H$\alpha$ luminosity above the median of the sample to those with a luminosity below the median. This method allows us  to eliminate the possibility that the trends observed in the scatter plots simply result from the fact that at higher redshifts larger volumes are sampled and therefore a larger number of high luminosity systems is to be expected; with our procedure the volume effect is removed.
In Figure   \ref{Lum_ha_z_AGN_frac} we  show that the strength of the H$\alpha$ line has indeed decreased over time. In the next section we will demonstrate that  this decrease is accompanied by a decrease in the radio luminosity of these active nuclei.  As noted earlier, this is part of a more general evolution of AGN strength in radio and X-rays observed from the  high- to the  low-redshift universe (\cite{Longair1966, Cowie2003}).

The trends observed are unlikely the result of the inclusion in the fiber, with increasing redshift,   of more H$\alpha$ emission from star-forming regions residing in  the outer part of the host: in order to explain the increase in  H$\alpha$ luminosity observed in each volume-limited sample, these star-forming regions would need to account for as much as three times the H$\alpha$ emission observed in the nuclear region at lower redshifts. This would severely affect the spectral  classification of these objects, making them fall in composite or SFG class.

\begin{figure}[h]
\begin{center}
\includegraphics[scale=0.4, angle=90]{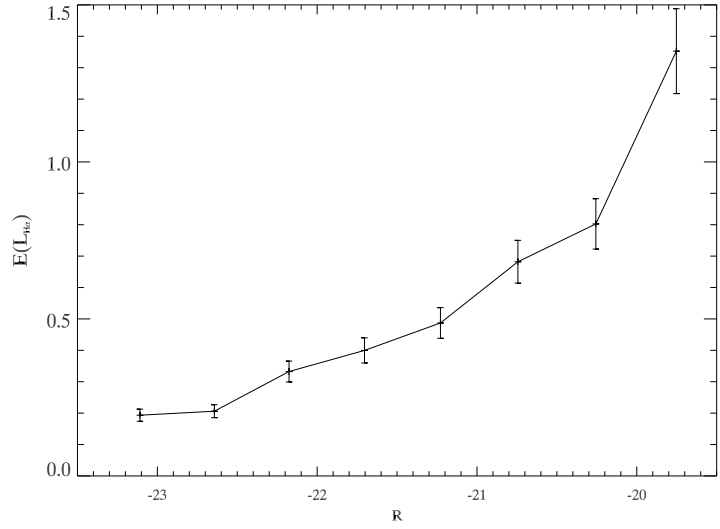}
\end{center}
\caption{\footnotesize \emph{The figure shows that in galaxies of different absolute magnitude $R$ the H$\alpha$ luminosity increases by a different factor E per Gyr (see text). AGN hosted by less luminous galaxies have seen their H$\alpha$ luminosities decrease by a larger factor in the past Gyr.}}
\bigskip
\label{rate_halpha_drop}
\end{figure}

Because of the requirement that each sample is  volume-limited, the look-back time is shorter for less luminous systems. In order to compare the various samples,  we have extrapolated the evolution in the look-back time considered to the same time interval $\Delta t=1$Gyr. In Figure \ref{rate_halpha_drop} we show the trend for the parameter  $E_{H\alpha}=log(L_{z4}/L_{z1})/(T_{z4}-T_{z1})$,  where $z4$ and $z1$ refer to the fourth (highest redshift) and first (lowest redshift) bin in each sub-sample and $T$ is the associated cosmic time in a standard $\Omega_m=0.3$, $\Omega_\Lambda=0.7$ cosmological model. This quantifies the order of magnitude by which the luminosity would increase over a 1 Gyr time interval, assuming a first-order  interpolation of the trends shown in Fig.\ref{Lum_ha_z_AGN}.Note that the bins for the lowest
luminosity galaxies are both sparsely populated {\it and} extend only
over a modest range of redshift, so the large changes seen may well be
exaggerated; over most of the luminosity range, the results suggest a
factor of 2-3 decrease in H$\alpha$ luminosity over the past gigayear.
 The evolution of the AGN H$\alpha$ power in less luminous galaxies is stronger  than the evolution of this quantity in  higher luminosity systems. This finding is in agreement with the evidence of downsizing in AGN activity, with AGN activity peaking at earlier epochs in more massive systems (\cite{Cowie2003, Kriek2007}).
 According to our  results, the most massive galaxies, have decreased their H$\alpha$ by circa two orders of magnitude in the past 10 Gyr, which means that at early times they had luminosities typical of strong AGN and quasars, a fact that we will further discuss in Paper III.   We note, however, that if black holes in less massive systems have been growing significantly only in recent epochs as suggested by downsizing, low-mass systems  have not been steadily decreasing their luminosity over the past $\sim 10$ Gyr like the most massive ones (which would make these low-mass systems surprisingly bright at very high redshifts) but only in a much more recent past.

\begin{figure}[t]
\begin{center}
\includegraphics[scale=0.5, angle=90]{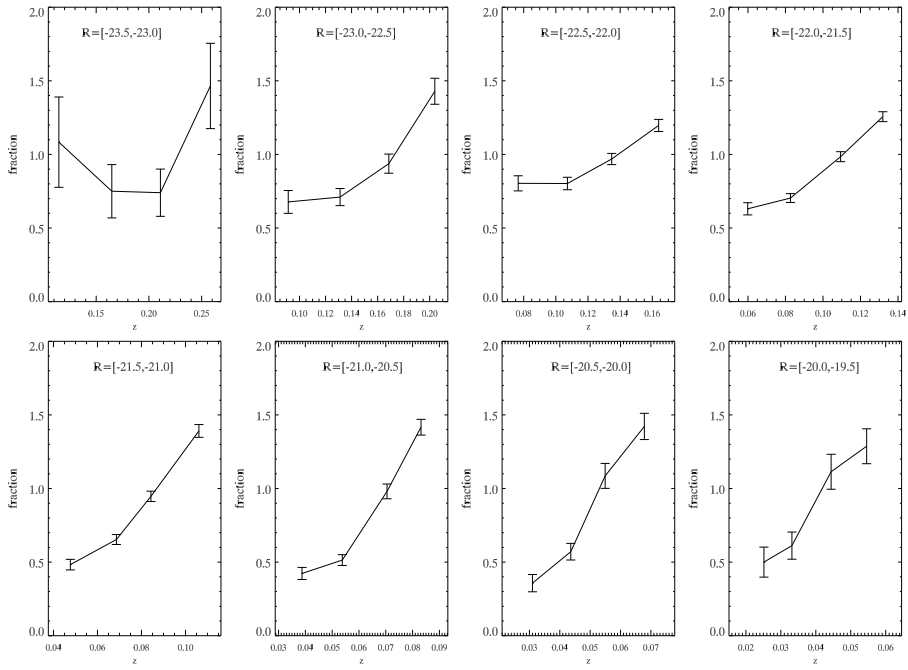}
\end{center}
\caption{\footnotesize \emph{The ratio of AGN host galaxies  with concentration parameter C above and below the median  in the eight volume-limited samples considered in this study. Galaxies harboring AGN evolve toward higher light concentrations  (lower C).}}
\bigskip
\label{conc_AGN_frac}
\end{figure}

\begin{figure}[h]
\begin{center}
\includegraphics[scale=0.5, angle=90]{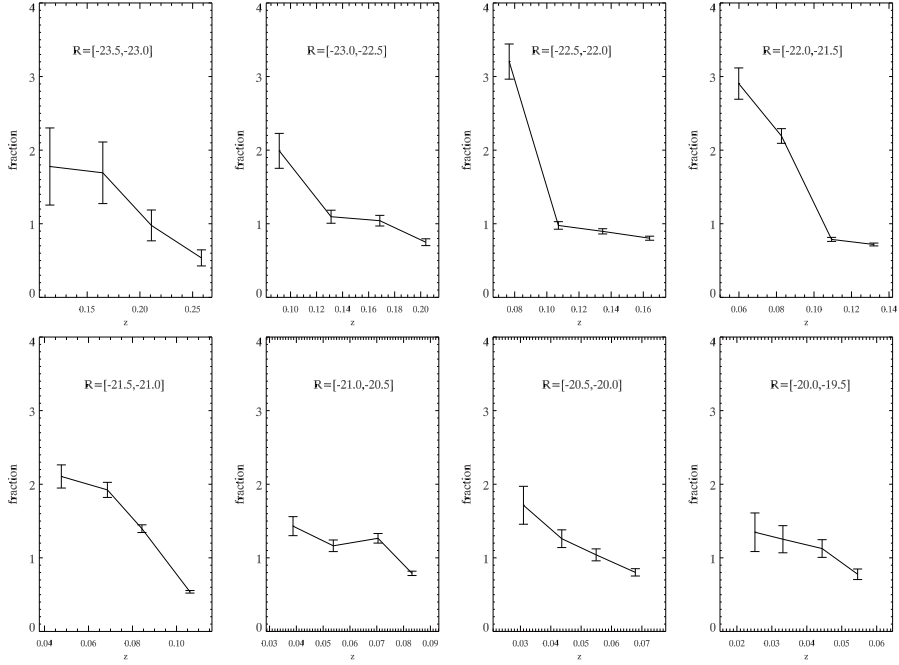}
\end{center}
\caption{\footnotesize \emph{Ratio of galaxies hosting spectroscopic AGN (corrected for dilution) with D$_{4000}$ parameter  above and below the median in the eight volume-limited samples.}}
\bigskip
\label{D4000_AGN_frac}
\end{figure}

\begin{figure}[h]
\begin{center}
\includegraphics[scale=0.4, angle=90]{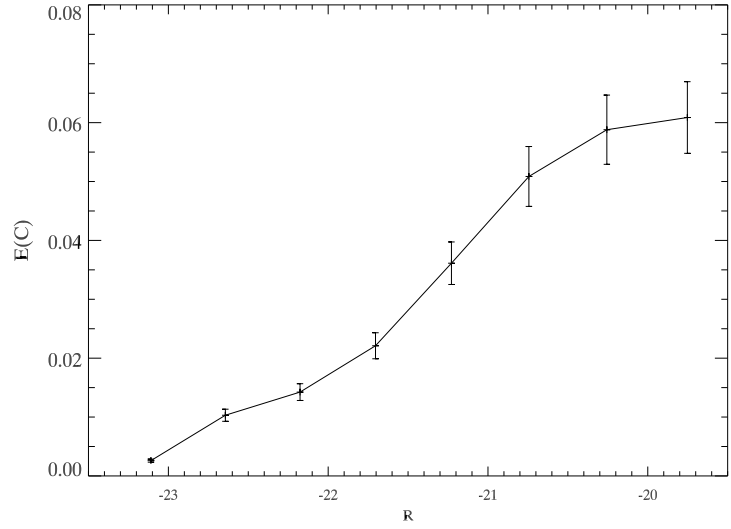}
\end{center}
\caption{\emph{The figure shows that in active galaxies of different absolute magnitude R the concentration parameter increases by a different factor E(C) per Gyr, looking back in time. Less luminous galaxies have decreased their concentration parameter  more in the past Gyr, a sign that their morphology has become more bulge-dominated.}}
\bigskip
\label{rate_conc_drop}
\end{figure}

Interestingly, this change in strength is accompanied by an increase in the concentration of the light as shown in figure \ref{conc_AGN_frac} and  an increase in Dn$_{4000}$ parameter describing the average time since the last burst of star formation (figure \ref{D4000_AGN_frac}). This suggests a co-evolution of star-formation, AGN and the morphological properties of galaxies over cosmic time.
The transformation of the light concentration is stronger in less luminous systems: this is shown in  figure \ref{rate_conc_drop} where we plot the trend for the parameter  $E_c=(C_{z4}-C_{z1})/(T_{z4}-T_{z1})$ which quantifies the factor by which  the concentration parameter C  increases over  1 Gyr, assuming a first-order  linear interpolation. We note that the C parameter spans a range in values between  0.6 for disks and  0.2  for ellipticals --- a change of $\sim$ 0.04 in such parameter is  therefore a $\sim$ 10\% effect.

This suggests that bulges in less massive systems are forming at recent epochs, in agreement with the findings of \cite{Cimatti2006}.

\begin{figure}[!h]
\begin{center}
\includegraphics[scale=0.5, angle=90]{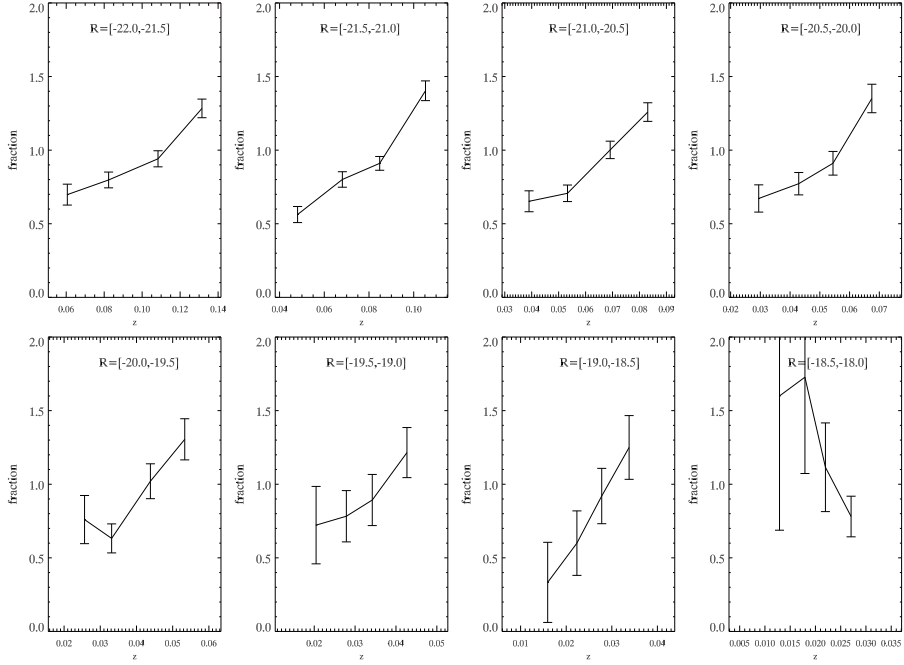}
\end{center}
\caption{\emph{Ratio of galaxies with composite spectra that have  concentration parameter C above and below the median  in  eight volume-limited samples spanning absolute magnitudes -22$<$R$<$-18. With the possible exception of the lowest luminosity bin, galaxies with composite (SFG+AGN) spectra evolve toward higher light concentrations (lower C).}}
\bigskip
\label{conc_COM_frac}
\end{figure}

\begin{figure}[!h]
\begin{center}
\includegraphics[scale=0.5, angle=90]{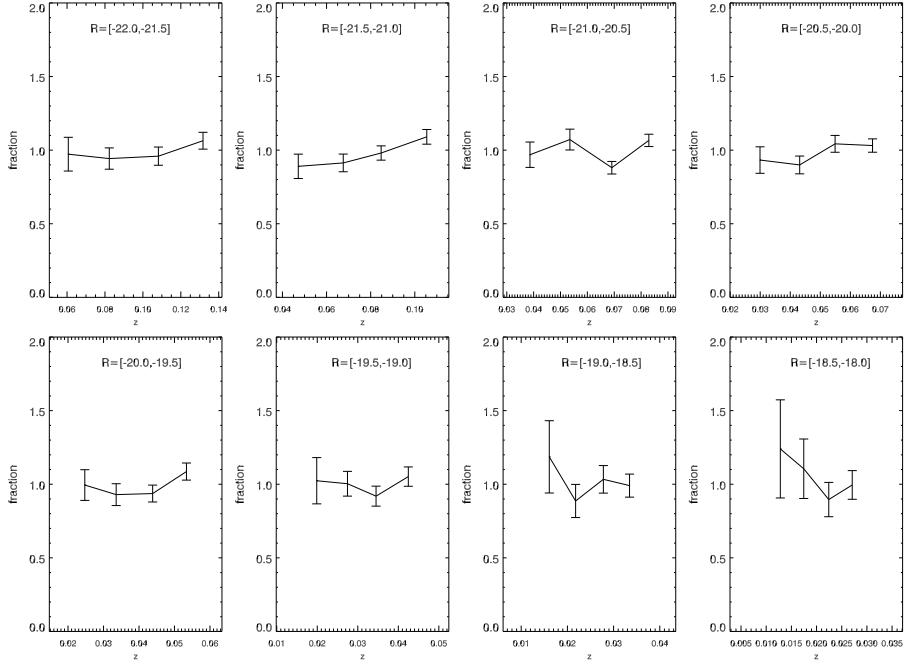}
\end{center}
\caption{\footnotesize \emph{Ratio of galaxies with star-forming spectra that have concentration parameter C above and below the median in  eight volume-limited samples spanning absolute magnitudes -22$<$R$<$-18. Galaxies with star-forming spectra do not show any sign of evolution over the look-back time considered.}}
\bigskip
\label{conc_SFG_frac}
\end{figure}

We further inspected the variation in the light-concentration for the other spectroscopic types; since star-forming  and composite systems are associated with lower luminosity systems, we have selected different intervals in magnitude: -22.0$<$R$<$-18.0.
While for composite systems we do observe a change in the number of systems with concentration above the median of each sample in their respective look-back times (fig.\ref{conc_COM_frac}), the same is not observed in SFG galaxies (Fig. \ref{conc_SFG_frac}). This is in agreement with the fact that AGN activity is strongly correlated with an enhancement in the concentration of galaxy light, typical of the presence of a bulge, and suggests that the disk/SFG phase is a relatively stable phase for galaxies, while the AGN/bulge phase is not: i.e., the transition from a disk to a bulge-dominated system must be rapid. From this point of view, the existence of  the red and blue sequences and a clean cut in morphology for galaxies at U-R =2.22 \cite[]{Strateva2001} can be interpreted as a consequence of the rapid evolution of systems of intermediate size, from star-forming disks to early-type systems, in agreement with the suggestions \cite{Balogh2004}.

\section{Evolution of AGN radio emission}
\label{aging_radio}

In Paper I we have shown that $\sim 5\%$ of emission-line AGN  at z $<$ 0.2 show radio emission  detectable by FIRST and/or NVSS surveys. The undetected emission line population has a mean radio flux density detectable by stacking FIRST images that is only an order of magnitude below the detection threshold.  Furthermore we have shown that there exists a  spectroscopically unremarkable population of AGN detected in FIRST which accounts for 4\% of the early-type population.

Since radio emission and optical emission lines  are independent features of an active nucleus,  with one not necessarily requiring the other \cite[]{Zirbel1995, Best2005, Reviglio2006},  we have explored the evolution of AGN activity in the SDSS look-back time for AGN radio emission, in order to see if the tail of the well-known decrease in  radio luminosity over cosmic time \cite[e.g.,][]{Longair1966} can be detected in the look-back time of the SDSS.

Using the technique described in the previous sections, we have selected a chain of volume-limited samples and we have plotted radio luminosity in each redshift range selected. Since NVSS and FIRST are both flux-limited surveys, selecting a volume-limited sample in the optical does not remove selection effects in the radio band: brighter radio sources are expected to be selected at higher redshifts in all volume-limited samples. In order to avoid this effect produced by the incompleteness of the radio sample at higher redshifts, we have used the  aforementioned stacking procedure \cite[]{White2007} to evaluate the median luminosity of radio AGN in bins of redshift for each volume-limited sample. The procedure effectively removes the selection given by the flux-limited nature of the radio surveys since all fields, including those  with source flux densities  below the survey flux-limit, are used for obtaining the median luminosity.

For each volume-limited sample we have thus considered four bins in redshift and plot the radio luminosity obtained by stacking all  fields containing emission-line AGN. The median luminosity was obtained by stacking in luminosity space and then fitting a two-dimensional gaussian profile to the median image obtained (see Paper I for details). 

\begin{figure}[h]
\begin{center}
\includegraphics[scale=0.5,angle=90]{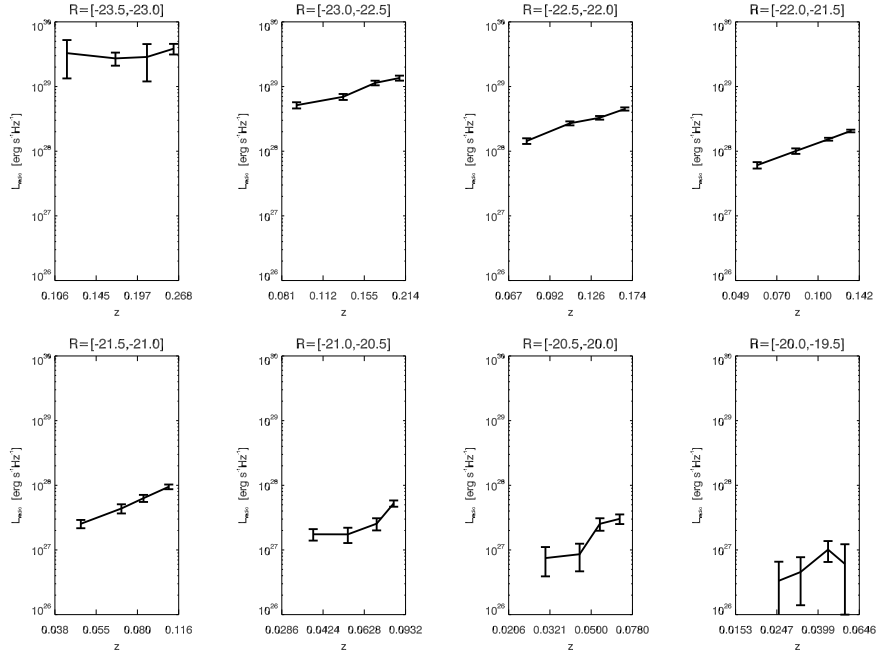}
\end{center}
\caption{\footnotesize \emph{The change in median radio luminosity of spectroscopically selected AGN  corrected for dilution in eight volume-limited samples, selected according to their absolute R magnitude. Error bars are evaluated from the first and third quartile images.}}
\bigskip
\label{stacks_evolution_AGN}
\end{figure}

The results are shown in figure \ref{stacks_evolution_AGN}.
We find that the average radio luminosity of emission-line systems has decreased over time in nearly all volume-limited samples. The average  K-correction  is of order 10\%,  too small to account for the decrease in radio luminosity found in the data. The stacked radio sources were unresolved (or marginally resolved): the radio luminosity is therefore equivalent to the peak radio emission coming from the central radio AGN. The fact that the FIRST survey resolves out part of the flux of extended sources therefore does not affect our analysis.
We note a relationship between the strength of the radio emission  and the optical luminosity of its host:  galaxies with lower optical luminosities harbor less intense radio AGN, a fact we will investigate further in Paper III.

The decrease in radio power is more pronounced in galaxies of lower luminosity, as we found for the H$\alpha$ emission:  as shown in figure \ref{stacks_evolution_AGN}, no evolution is seen in either the -23.5$<$R$<$-23.0 bin or the -20.0$<$R$<$-19.5: a lack of evolution in the concentration parameter  was also noted for the highest luminosity bin (fig. \ref{conc_AGN_frac}). In the -20.0$<$R$<$-19.5 bin the look-back-time may be too short for observing an actual change.

\begin{figure}
\begin{center}
\includegraphics[scale=0.5,angle=90]{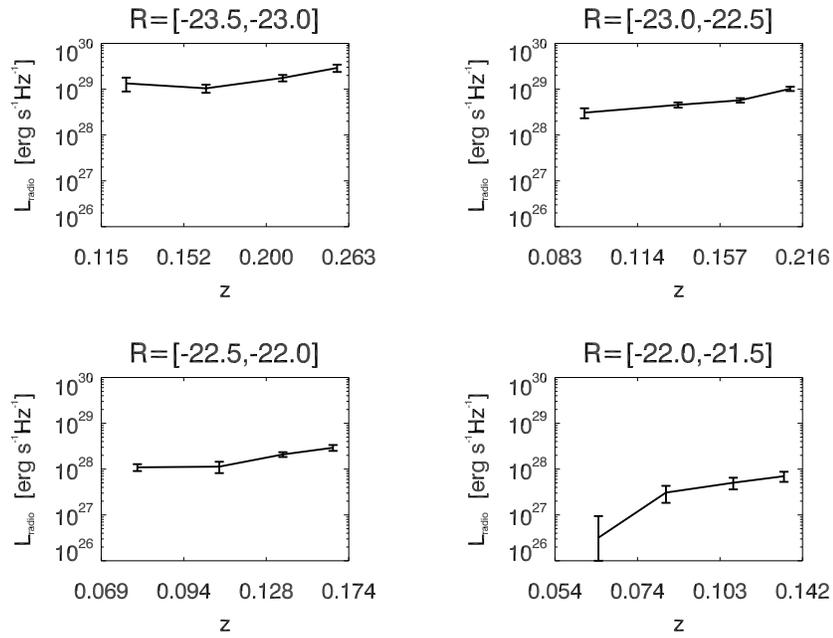}
\end{center}
\caption{\footnotesize \emph{The change in radio-AGN luminosity in spectroscopically passive galaxies in four volume-limited samples, selected according to their absolute R magnitude}}
\bigskip
\label{stacks_evolution_ABG}
\end{figure}

If we consider spectroscopically passive galaxies which show no significant signs of AGN activity in their spectra, we find a less marked decrease in the median power of their AGN from higher to lower redshift. This is shown in Figure  \ref{stacks_evolution_ABG}. For absolute magnitudes fainter than  -21.5, no flux is detected in our stacks ($<$ 10 $\mu$Jy), showing that low-luminosity hosts lacking emission lines harbor fainter, or non-existent,  radio AGN. In current radio AGN production models, the radio power is connected with accretion rate \cite[]{Meier2001}. If these systems have used up most of their gaseous reservoir at earlier times as suggested by the lack of emission lines, it is likely that their accretion rates are lower and their radio emission weaker. We will  investigate this fully in Paper III.

In summary, it appears that  more evolved, more luminous galaxies show less sign of evolution in radio emission over the  past 1 Gyr, consistent with the trends  found for the emission lines  and the morphology of the host.

\section{Evolution of AGN spectroscopic activity}
\label{aging_spectro}

In Paper I we showed that systems with unremarkable lines are often weak AGN with weak emission lines that  are misclassified as passive galaxies because of nuclear  light dilution by the host galaxy stellar light. 
We have shown that this population is on average redder than the population with detected lines, and  suggested that there is a physical difference between the two populations.

\begin{figure}[h]
\begin{center}
\includegraphics[scale=0.4,angle=90]{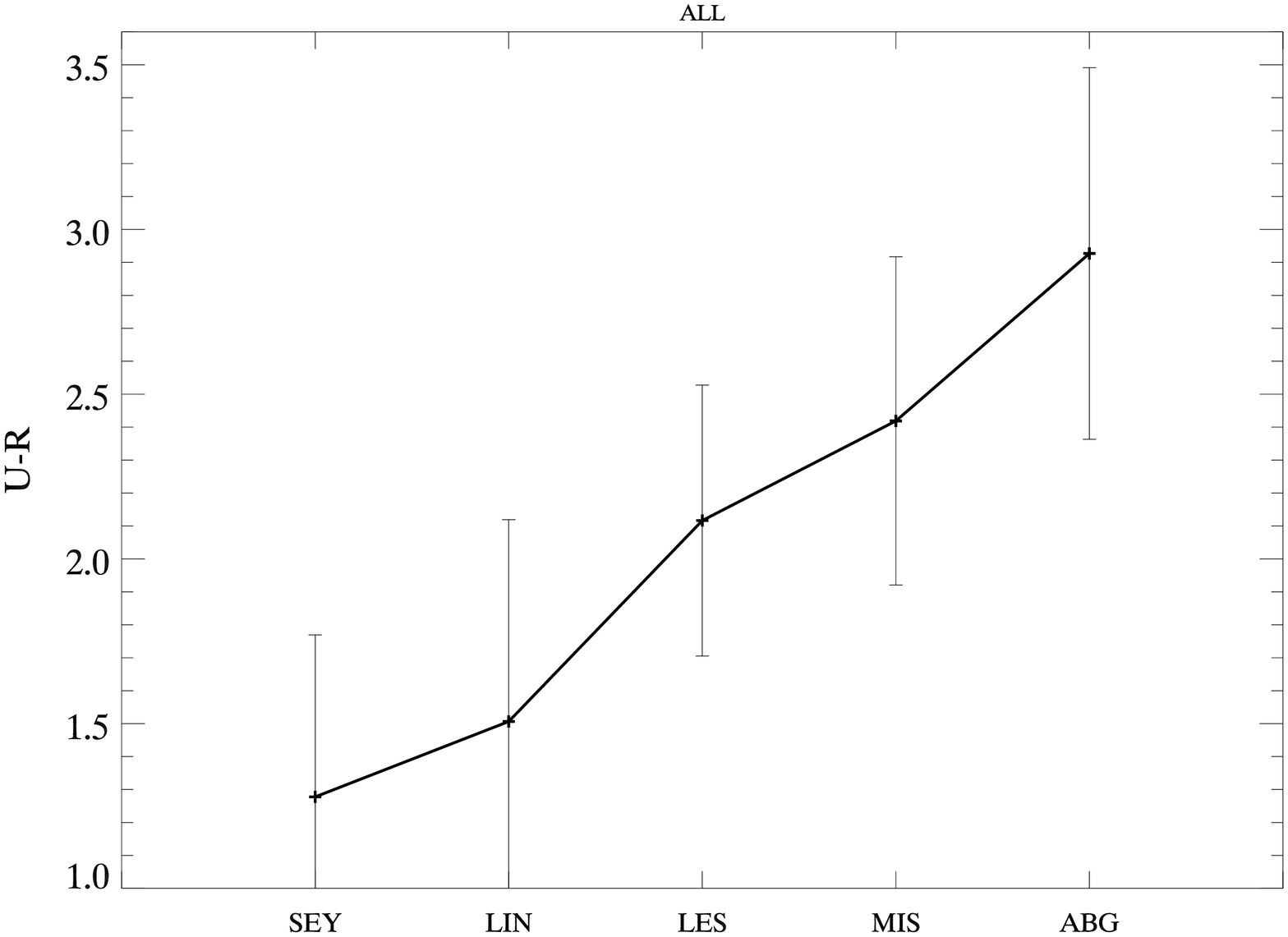}
\end{center}
\caption{\emph{Trend in the host color for different classes of AGN with increasingly unremarkable lines: Seyfert (SEY), Liners (LIN), Low-excitation systems  (LES) with marginally detected or undetected [OIII] or H$\beta$, misclassified LES systems (MIS) and truly passive galaxies with the X-ray or radio signature of an AGN (ABG). The bars represent the median absolute deviation in color for each class. }}
\label{color_trend_AGN_new}
\end{figure}

In Figure \ref{color_trend_AGN_new} we show that this is part of a more general trend of line diminution in redder systems. If we select all systems at z$<$0.2 with the four standard BTP lines detected (requiring that their EW is $>$1.0) and divide these systems between Seyfert II galaxies and Liners using the usual cut [OIII]/H$\beta >$3, we find that the Seyfert II population is on average bluer than the Liner population. If we consider the class of AGN with H$\alpha$ and [NII] detected (EW $>$1), but [OIII] and H$\beta$ marginally detected or undetected (EW$<$1), which we will  refer to as  LES (low excitation systems), we find that  this class is redder than the Liners.  If we then consider  the class with all four BTP lines marginally detected or not detected at all (EW of all four lines $<$1) which we reclassified as AGN according to the criteria discussed in Paper I, this class is redder than the LES class. Finally, if we consider the sub-sample of  truly spectroscopically passive galaxies with all lines not detected, but showing X-ray or radio AGN, we find that these galaxies are redder than all other classes. This strongly suggests that the line strength in active galaxies is strongly linked with the galaxy color.

The colors adopted have been corrected for dust reddening by means of the estimate of the extinction coefficient A$_z$ obtained by \cite{Kauffmann2003}.  
Estimate of the extinction in the U and R bands from the A$_z$ coefficient have been obtained by means of the extinction law calculated by \cite{Rieke1985}, where the  A$_z$/A$_v$ value needed for the conversion has been obtained by interpolating the values for the I and J band, since the average
wavelength of the Z band is 8931 Angstrom, in between the values for the standard  J and the I bands. The trends found are therefore not an effect of dust extinction, i.e., they are not given by a correlation between the nuclear extinction and the global extinction in the galaxy that would imply that more dusty galaxies also have  more dusty nuclei.

\begin{figure}[h]
\begin{center}
\includegraphics[scale=0.5,angle=90]{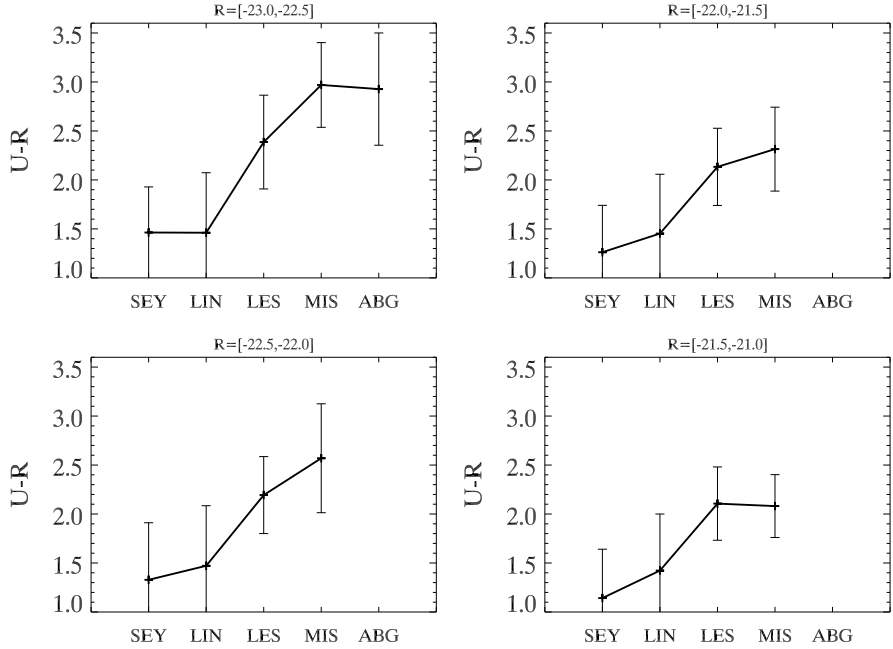}
\end{center}
\caption{\footnotesize \emph{Same as figure \ref{color_trend_AGN_new}, but for volume-limited sub-samples with narrow host luminosity ranges in R-band (and therefore small mass ranges in the supermassive black hole powering the AGN). At all luminosities, galaxies show less remarkable lines for redder colors, suggesting that the trend is not produced by the difference in the supermassive black hole mass, but more likely by the accretion properties of the central engine. When split into these several sub-samples, the ABG class with AGN confirmed by radio or X-ray emission in each sample becomes quite small; if fewer than 3 galaxies were present we did not considered the value statistically reliable and did not plot it.}}
\bigskip
\label{color_trend_AGN_split_new}
\end{figure}

In Figure \ref{color_trend_AGN_split_new} we show that the same trends are found when hosts with similar luminosities (and therefore with similar supermassive black hole mass) are chosen by restricting the sample to volume-limited sub-samples spanning intervals of only 0.5 in absolute $R$ magnitude. This suggests that the mass of the supermassive black hole is unlikely to be the parameter that drives the difference in the prominence of the lines.  As we will fully explore in Paper III, the accretion rate onto the supermassive black hole could be the parameter that regulates the prominence of the lines. It is reasonable to expect that, as systems become more and more gas depleted and therefore redder as a consequence of their decrease in star-formation rate, their accretion rates drop. Lower accretion rates would power less luminous nuclear emission, leading to less prominent recombination lines.

If galaxies progressively use up their gaseous reservoir, then more gas-depleted galaxies should have quenched their star formation at earlier times. In this case, the prominence of their lines in spectra should be correlated not only with the color of the galaxy, but also with the Dn$_{4000}$ \cite[]{Bruzual1983}and the H$_{\delta A}$ \cite[]{Worthey1997} parameters, which are indicators of the age of their stellar population.\footnote{ Dn$_{4000}$ quantifies the break in galaxy spectra occurring at 4000 \AA, a feature which  arises because of the accumulation of a large number of absorption lines in a narrow wavelength range. These absorption lines arise in stellar atmospheres. Young stars, having highly ionized atmospheres, are ineffective at absorbing photons, so galaxies with young stellar population have  weaker Dn$_{4000}$ breaks. 
H$\delta_A$, on the contrary, quantifies the strength of the H$_{\delta A}$ absorption  line, which is typical of A stars.   H$_{\delta A}$ therefore peaks in galaxies in  which the light is dominated by such a population and then decreases as the A-star population fades and the galaxy becomes redder.}

\begin{figure}[t]
\begin{center}
\includegraphics[scale=0.4,angle=90]{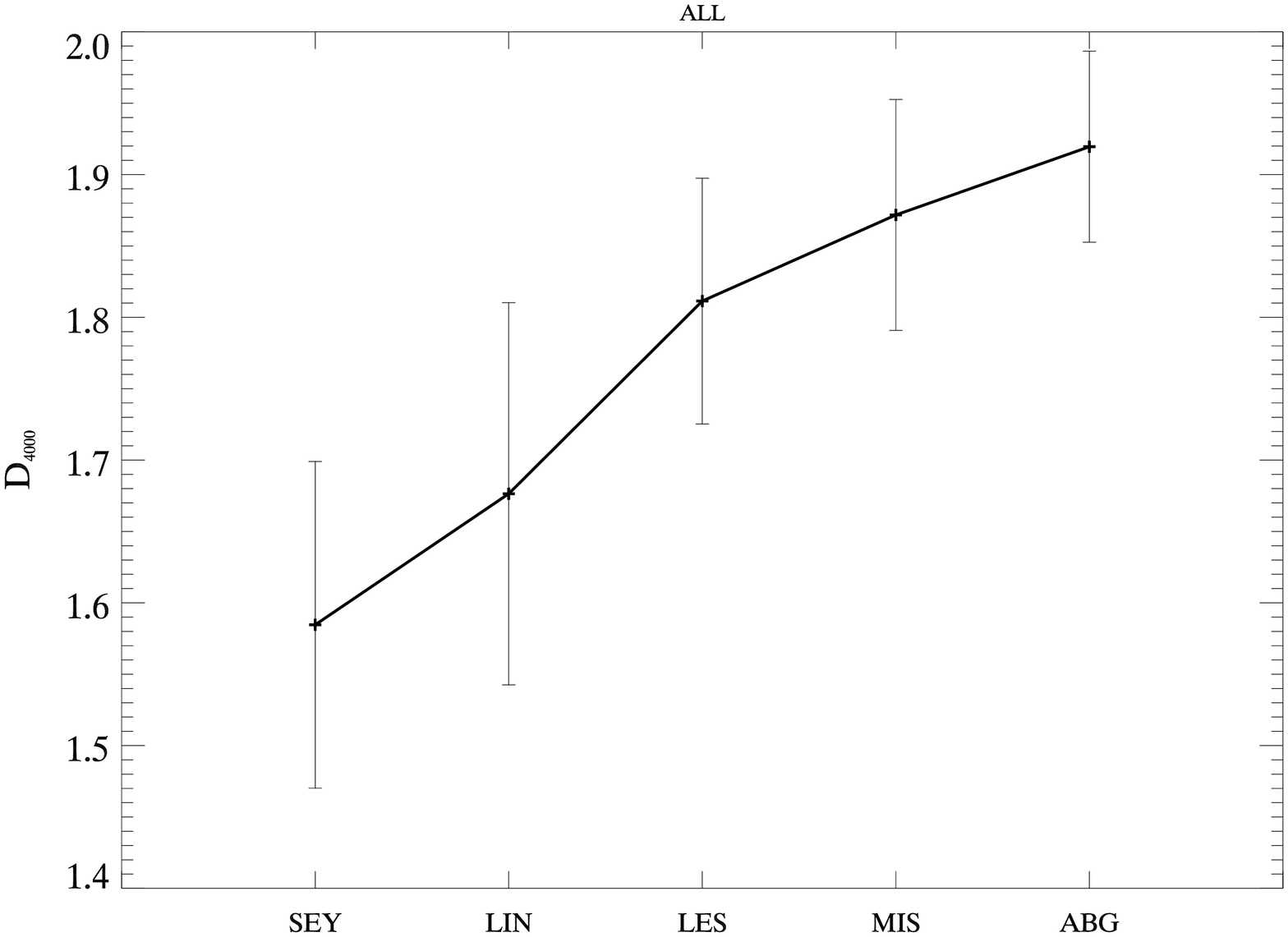}
\end{center}
\caption{\footnotesize \emph{Trend in D$_{n4000}$ break strength for different classes of AGN with increasingly unremarkable lines: Seyfert (SEY), Liners (LIN), Low-excitation systems  (LES) with marginally detected or undetected [OIII] or H$\beta$, misclassified LES systems (MIS) and spectroscopically passive galaxies with the X-ray or radio signature of an AGN (ABG). The bars represent the median absolute deviation in color for each class. }}
\bigskip
\label{d4000_trend_AGN}
\end{figure}

\begin{figure}[!t]
\begin{center}
\includegraphics[scale=0.4,angle=90]{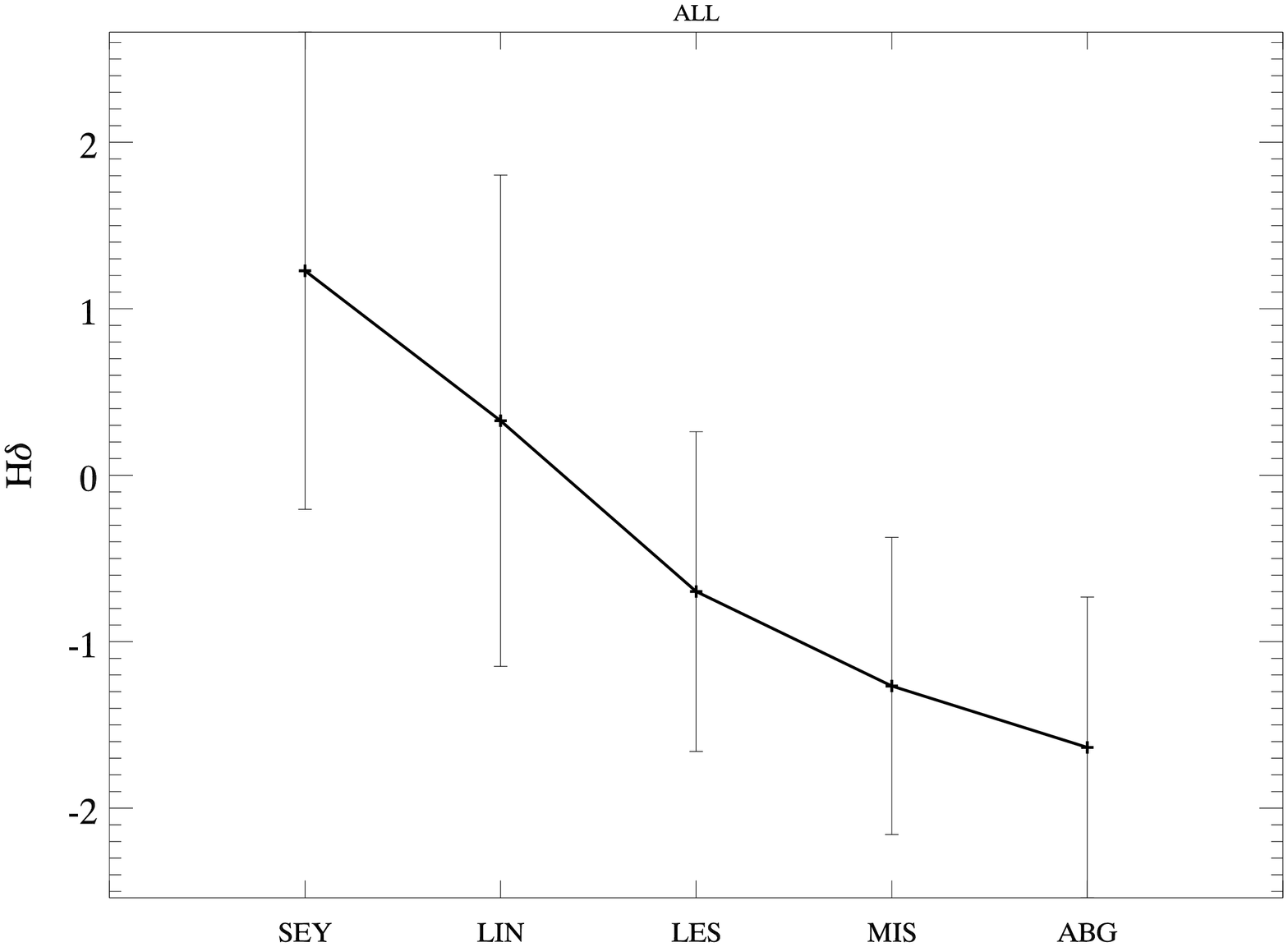}
\end{center}
\caption{\footnotesize \emph{Trend in $H_{\delta A}$  for different classes of AGN with increasingly unremarkable lines: Seyfert (SEY), Liners (LIN), Low-excitation systems  (LES) with marginally detected or undetected [OIII] or H$\beta$, misclassified LES systems (MIS) and spectroscopically passive galaxies with the X-ray or radio signature of an AGN (ABG). The bars represent the median absolute deviation in color for each class.}}
\bigskip
\label{Hdelta_trend_AGN}
\end{figure}

\begin{figure}[!b]
\begin{center}
\includegraphics[scale=0.5,angle=90]{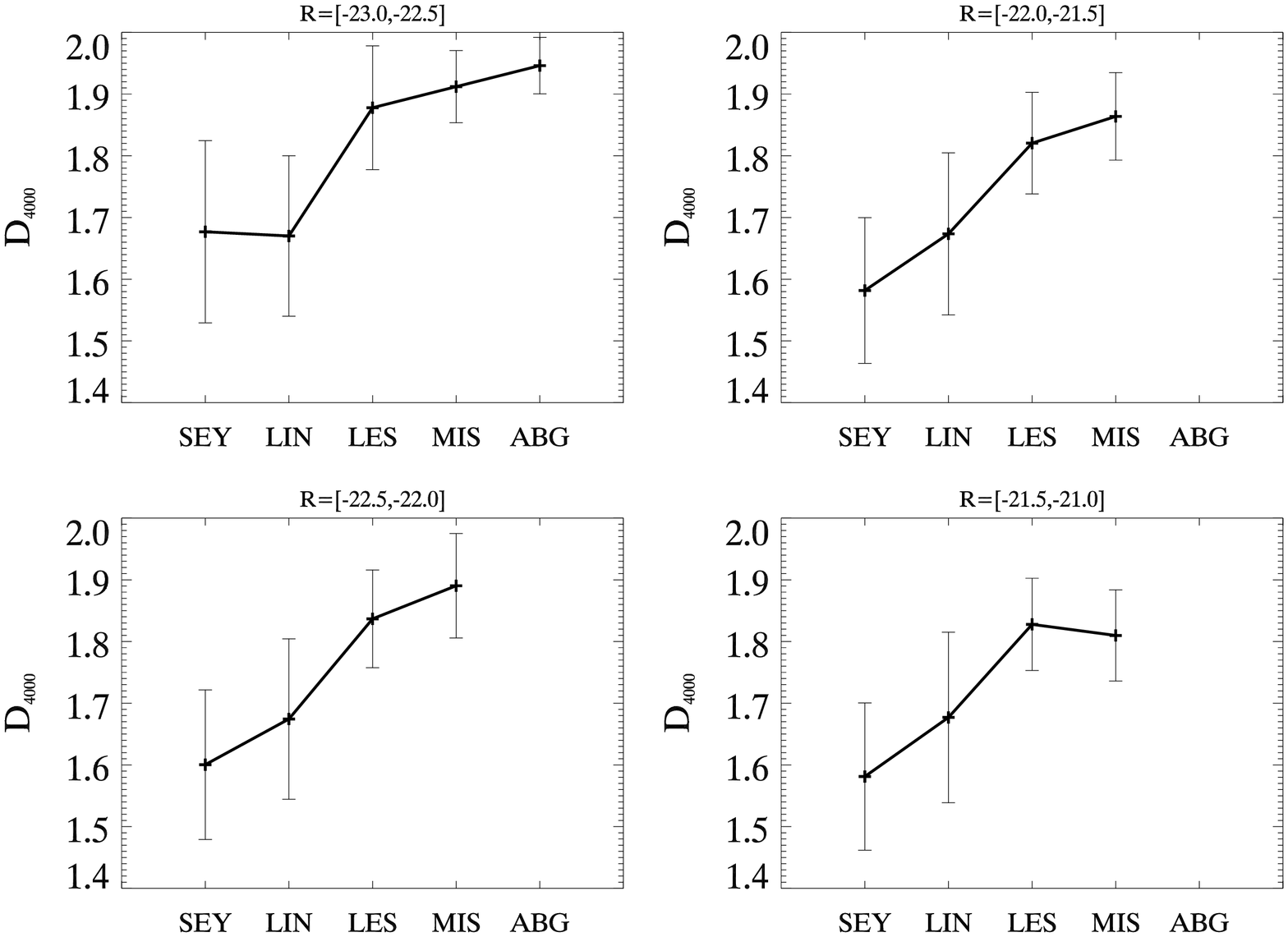}
\end{center}
\caption{\footnotesize \emph{Same as figure \ref{d4000_trend_AGN}, but for volume-limited sub-samples with a narrow range of host luminosity in R-band (and therefore in the mass of the supermassive black hole powering the AGN). At all luminosities, galaxies which show less remarkable lines  also show stronger D4000 breaks, showing that systems with less remarkable lines are hosted by systems with older stellar populations, and therefore are more gas depleted.}}
\bigskip
\label{d4000_trend_AGN_split}
\end{figure}

\begin{figure}[h]
\begin{center}
\includegraphics[scale=0.5,angle=90]{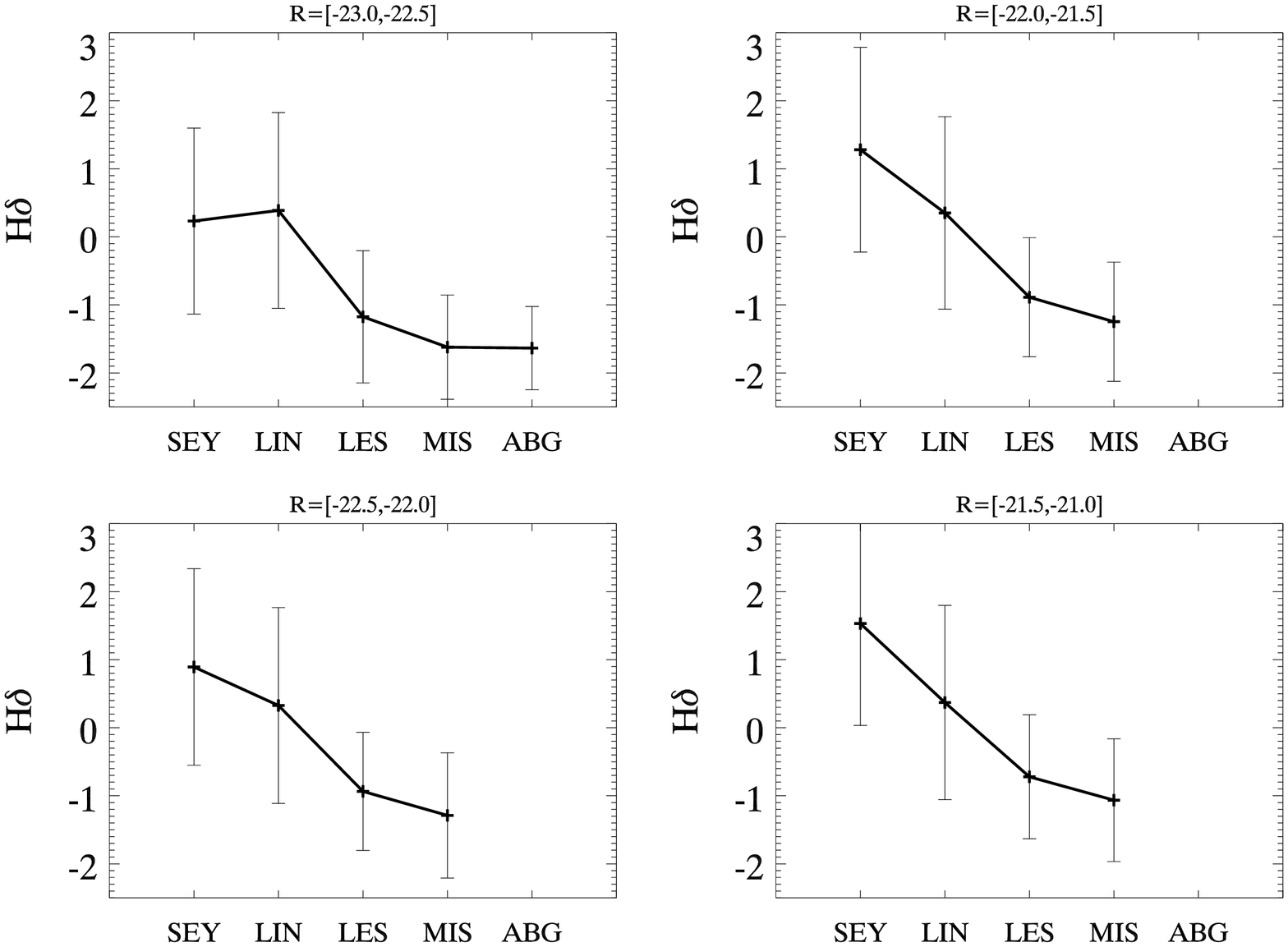}
\end{center}
\caption{\footnotesize \emph{Same as figure \ref{Hdelta_trend_AGN}, but for volume-limited sub-samples with a narrow range of  host luminosity in R-band (and therefore mass of the supermassive black hole powering the AGN). At all luminosities, galaxies which show less remarkable lines  also show lower $H_{\delta A}$, showing that systems with less remarkable lines are hosted by systems with older stellar populations, and therefore are more gas depleted.}}
\bigskip
\label{Hdelta_trend_AGN_split}
\end{figure}

In Figure \ref{d4000_trend_AGN} and \ref{Hdelta_trend_AGN} we show the median values of the Dn$_{4000}$ and H$_{\delta A}$ parameters for the five different classes of objects discussed above. It is clear that weaker lines are associated with galaxies with older stellar populations, since the trend for Dn$_{4000}$ is decreasing and the trend for H$_{\delta}$ is increasing.
If we split our sample into four classes of luminosity as before, we find that such trends persist for all types of galaxies confirming that, at all galaxy luminosities, lines become weaker as the star-formation is quenched and the galaxy stellar population becomes redder (Fig.  \ref{d4000_trend_AGN_split} and \ref{Hdelta_trend_AGN_split}).

\begin{figure}[!h]
\begin{center}
\includegraphics[scale=0.5,angle=0]{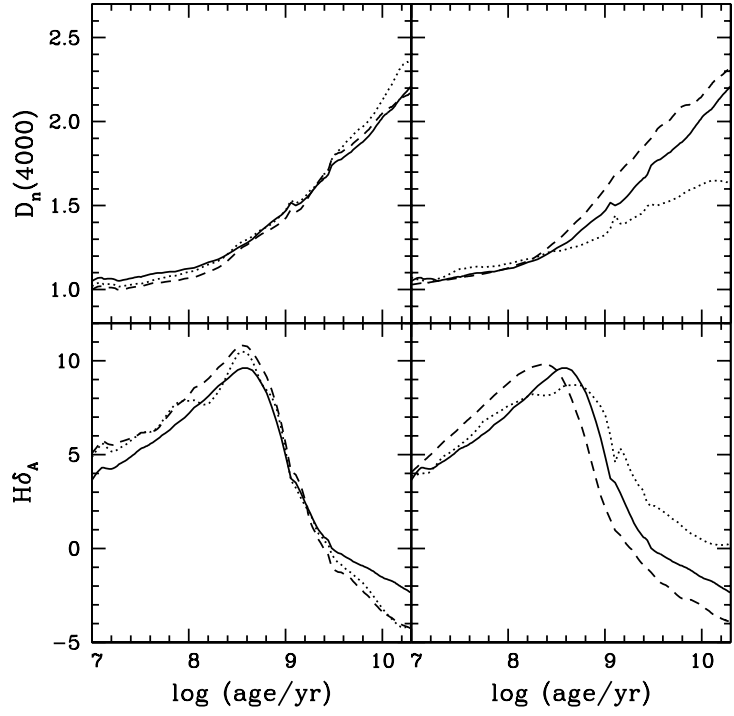}
\end{center}
\caption{\footnotesize \emph{A plot showing the evolution with time of the D$_n$(4000) parameter after a burst of star-formation, from \cite{Kauffmann2003b}}}
\bigskip
\label{kauff_D4000_time}
\end{figure}

If we consider the evolution of the Dn$_{4000}$ parameter after a burst of star-formation as shown in Kauffmann et al. 2003 (we show their plot in Fig. \ref{kauff_D4000_time})  we find that a variation from 1.6 to 1.9 in the D4000 parameter corresponds approximately to  an interval of 4$\times 10^9$ years.  At each luminosity, the ABG galaxies had their last burst of star formation $\sim  4\times 10^9$  years earlier than the Seyfert population.
If we consider Seyferts and Liners we have a variation of  Dn$_{4000}$ of $\sim 0.05$, which corresponds to $ \sim 4\times10^8$ yrs. This implies that on timescales of $4 \times 10^8$ yrs a significant variation of the nuclear spectrum can be observed.

The look back time for the SDSS survey is $\sim 2\times 10^9$. Therefore, if this line of reasoning is correct, a decline in the strength of lines for a fixed host should be observable, as shown in section \ref{evolution}.

\section{Environmental dependency of activity}
\label{environment}

Differences in star-formation and AGN activity in galaxies may depend on  their intrinsic properties, such as the depth of their potential well or their  richness in gas, or by the effects of the environment in which they reside. Denser environment favor close encounters among galaxies which are known to disturb the distribution of their interstellar medium, triggering star-formation activity and possibly AGN activity. We have therefore explored further  the dependence of activity and morphology on environment and galaxy mass/luminosity to  clarify this issue.

We have associated all galaxies with a density calculated by sampling the volume defined by the tenth nearest neighbor to a galaxy and statistically correcting for unseen galaxies, as described in \cite{Carter01} and in Paper I. This provides us  with an estimate of the environmental density on a few megaparsec scale, effectively dividing the sample into the different regions of large-scale structure: voids, filaments, and clusters. The estimator chosen is not sensitive to substructure on smaller scales and therefore does not ``see'' cluster cores or tight pairs, the study of which is beyond the scope of this paper.

\begin{figure}[h]
\begin{center}
\includegraphics[scale=0.6,angle=0]{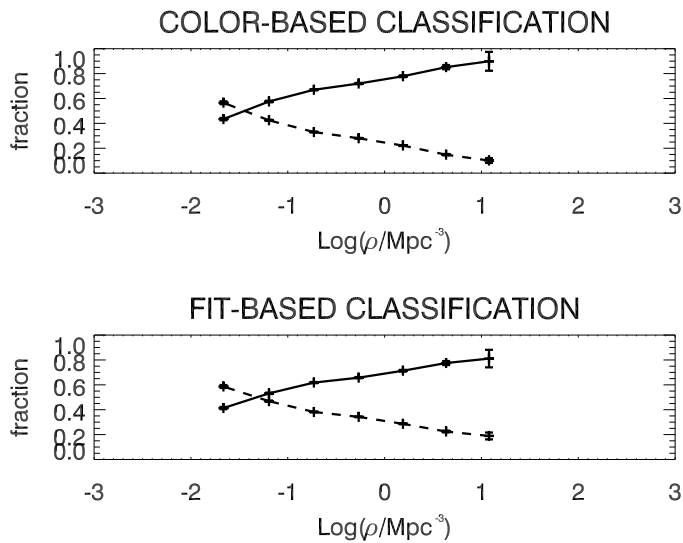}
\end{center}
\caption{\footnotesize \emph{The density-morphology relation in our sample. Solid lines trace the fraction of early type galaxies, while dashed lines trace the fraction of late-type systems. In the upper plot galaxies are divided into early and late-type based on their u-r color (with  u-r$>$2.22 for early-type). In the lower plot, galaxies are classified as early-type if the likelihood of a de-Vaucouleur profile for their surface brightness is higher than the likelihood of an exponential fir and late-type otherwise. }}
\bigskip
\label{density_morphology}
\end{figure}

In figure \ref{density_morphology} we show that  the well-known  density-morphology relation, originally observed in clusters by \cite{Dressler1980}, is observed also in the lower density regime of filaments and voids, with an increase in late type systems with decreasing large-scale environmental density. This trend is observed both  both if we divide early- and late-types according to color following \cite[]{Strateva2001} or on the basis of the likelihood of an exponential or de-Vaucouleurs fit to their light profiles.

This change in the typical morphology of galaxies with increasing density   is accompanied by a change in activity as well.   In Figure \ref{fraction_rho_final}  we show the fractional abundance as a function of density for the four main spectroscopic types, after correction for dilution as discussed in Paper I: star-forming galaxies (SFG), AGN, composite galaxies (star-forming plus AGN) and passive galaxies showing only absorption lines in their spectra (ABG).
It is evident how galaxies with different types of activity inhabit different environments: star-forming galaxies preferentially reside in low-density environments, galaxies showing both star-formation and AGN activity signatures  in environments of intermediate density, galaxies with spectroscopic AGN activity  in dense environments and finally, galaxies lacking emission lines in very dense environments.
As discussed in Paper I, 4\% of these passive galaxies do harbor radio AGN at the sensitivity threshold of FIRST and NVSS, and the majority of such galaxies host radio AGN typically one order of magnitude fainter. The ABG class, therefore, should be considered as passive only spectroscopically.

\begin{figure}[h]
\begin{center}
\includegraphics[scale=0.5]{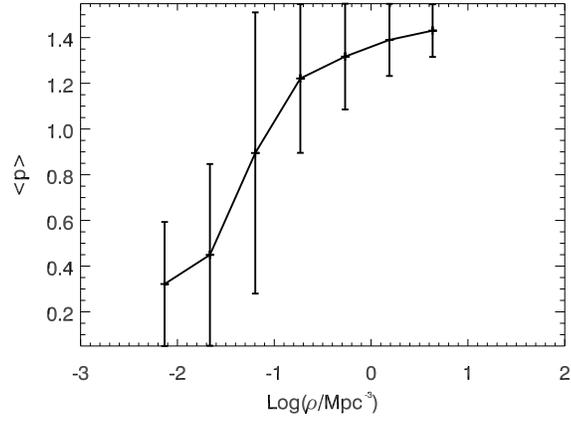}
\end{center}
\caption{\footnotesize \emph{Trend in the median bulginess of the systems with varying environmental density as traced by the median $p$ parameter. High values of p are associated with bulge-dominated systems. Error bars represent the median absolute deviation of $p$ in each density bin. }}
\bigskip
\label{bulgy_density}
\end{figure}

 We note that the smooth  transformation of activity with  environment from star-formation-dominated to AGN-dominated and eventually into galaxies lacking signs of nuclear recombination lines  is not only accompanied  by a change in  the relative abundance of early- to late-type galaxies in denser environments (as described by the usual density-morphology relation), but also by a smooth transition in the morphology of galaxies, with an increasing  prominence of the bulge relative to the disk. In order to quantify this variation we have considered two different indicators. The first takes advantage of the fact that  the light profile of disk galaxies is  well-described by an exponential law, while the light profile of  bulges is well-described by a de-Vaucouleurs law. As a consequence, for each galaxy, the parameter $p$ defined as the arcotangent of the ratio of the exponential fit likelihood $L_{exp}$ to the de-Vaucouleurs fit likelihood $L_{deV}$, $p=atan(L_{exp}/L_{dev})$, is a measure of the probability that the system is disk or bulge dominated. Values of $p$ close to $\pi/2$ describe galaxies with  a very high probability of being bulge-dominated, while values of $p$ close to zero describe galaxies with a very high probability of being disk-dominated. Values in between represent systems with intermediate features or irregular galaxies.  The smooth variation from disk-dominated systems to bulge-dominated systems with environment is shown in figure \ref{bulgy_density}.

\begin{figure}[t]
\begin{center}
\includegraphics[scale=0.5]{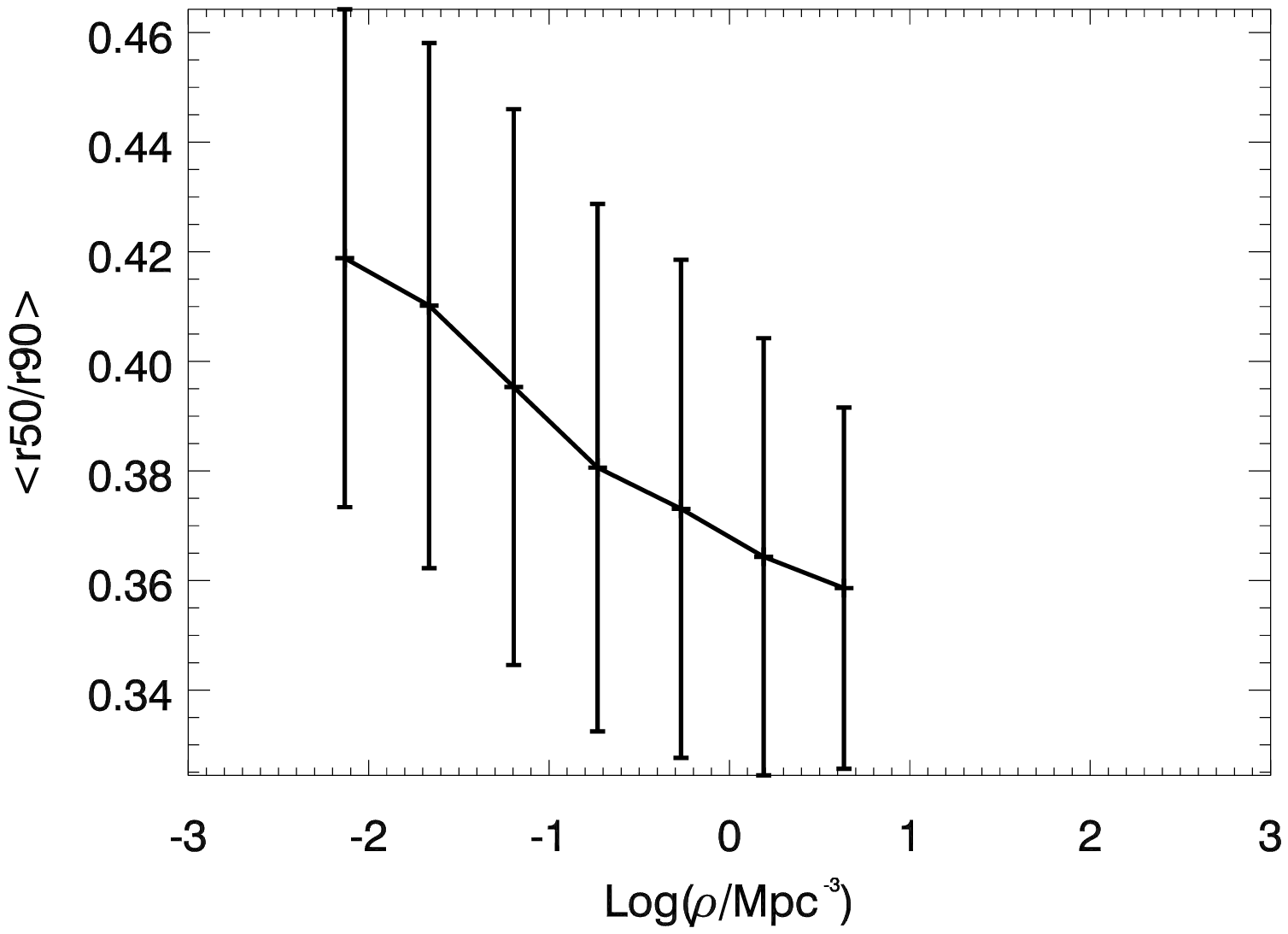}
\end{center}
\caption{\footnotesize \emph{Trend in the median bulginess of the systems with varying environmental density as traced by the median concentration parameter $C$. High values of $C=r50/r90$  are associated with disky systems.  Error bars represent the median absolute deviation of $C$ in each density bin.}}
\bigskip
\label{conc_density}
\end{figure}

A transformation from disk- to bulge-dominated systems along the Hubble sequence can also be traced using a different estimator --  the concentration parameter C, given by the ratio of the radius containing 50\% of the galaxy light to the radius containing 90\% of the light. Moving from low to high values of C signifies a change in the concentration of the light which describes a change in morphology from bulge-dominated to disk-dominated systems, since the parameter is inversely proportional to the concentration of the light.  The smooth variation of the value of C  with environment is shown in figure \ref{conc_density}. This shows that the correlation of density with morphology is not simply given by a different fraction of  ellipticals to spirals in different environments, but that the bulginess of systems varies smoothly across environments. This change accompanies the variation in the type of activity with environment noted above.

\begin{figure}[h]
\begin{center}
\includegraphics[scale=0.5]{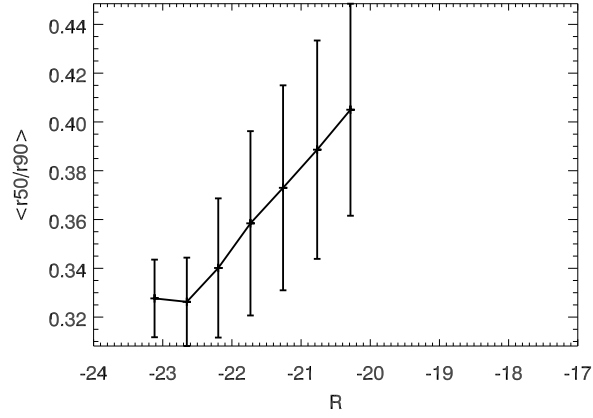}
\end{center}
\caption{\footnotesize \emph{Trend in the median bulginess of the systems with absolute $R$ magnitude. More luminous systems are on average more bulgy, a well-known feature of the Hubble Sequence.  Error bars represent the median absolute deviation of $C$ in each bin of magnitude.}}
\bigskip
\label{conc_abs_R}
\end{figure}

Since the average mass and luminosity of galaxies increases with increasing  environmental density and the prominence of bulges increases with increasing  luminosity/mass as shown in Figure \ref{conc_abs_R}, the trends observed may result from either environmental effects \cite[]{Park2007}  or the different evolution of galaxies of different mass \cite[]{Balogh2004}.

\begin{figure}[h]
\begin{center}
\includegraphics[scale=0.5, angle=90]{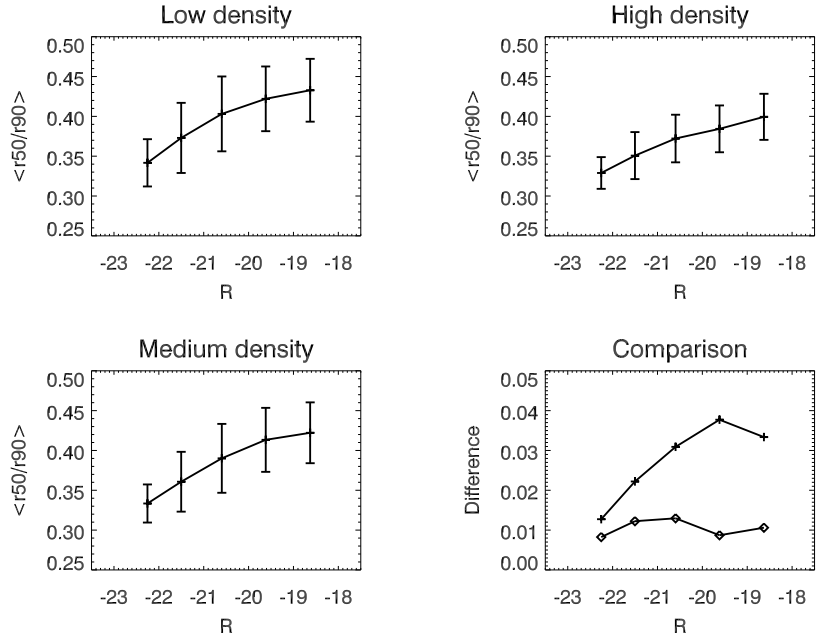}
\end{center}
\caption{\footnotesize \emph{The correlation of light concentration with the luminosity of a galaxy defining the Hubble Sequence is independent of the density of the environment -- very similar trends  are found in all  environments. The contribution of the environment appears to be limited to a small, yet significant, offset of such trends toward more concentrated systems in denser environments. This is show in the lower-right plot, where crosses trace the difference in the median values of the high- and the low-density samples and the diamonds the difference in the median values of the medium-  and the low-density samples.}}
\bigskip
\label{morpho_lum}
\end{figure}

In order to explore further this degeneracy,  we compared the actual effect of environment on the morphology of galaxies,  selecting three environments (low density $-2.0<log(\rho)<-1.0$, medium density $-1.0<log(\rho)<0.0$ and high density $log(\rho)>0.0$) and evaluating the average bulginess of systems with different luminosities in each environment. If the environment plays a crucial role in shaping the properties of galaxies, we would  expect a significant difference in the trends across different environments: all galaxies in denser environments must have undergone more interactions than galaxies in low density environments and at all luminosities they should be significantly more evolved towards earlier types.  The results are shown in Figure \ref{morpho_lum}.
We find that a similar correlation between luminosity and bulge prominence is in place  in high-, medium-, and  low- density environments: higher luminosity systems show increased bulginess and the trends are quite smooth. This is in agreement with the claim by \cite{Hogg2004} that the color-magnitude relation for galaxies does not depend significantly on the environment and   suggests that whatever  mechanism transforms galaxies into bulgier systems  must be effective in all types  of environments from low to high density. From this point of view, the environment seems to play a lesser role in shaping the morphology of galaxies;  the magnitude of its effect can be evaluated by comparing the offsets in the trends of Figure \ref{morpho_lum} as shown in the lower right corner of Figure \ref{morpho_lum}. This offset accounts for just a few percent variation: the bulk of the correlation must be produced by a mechanism that works equally well in all environments.

\begin{figure}[t]
\begin{center}
\includegraphics[scale=0.5,angle=0]{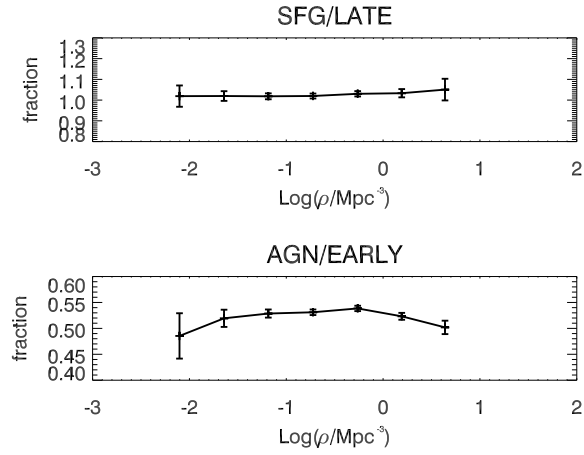}
\end{center}
\caption{\footnotesize \emph{Beyond the density-morphology correlation: the fractional abundance of SFG and AGN is not correlated with the environment if we normalize AGN to the number of early-type galaxies and SFG to the number of late-type galaxies. Fractions higher than 1.0 for the SFG/LATE plot are found because of the existence of early-type systems spectroscopically classified as star-forming.}}
\bigskip
\label{sfg_late_agn_early}
\end{figure}

We have conducted a similar analysis on star-forming galaxies  and AGN hosts in order to evaluate whether or not the dependence of activity on density is primarily an effect of the density-morphology correlation. In Figure \ref{sfg_late_agn_early} we show that the fraction of AGN among early-type galaxies in different environments is remarkably flat, suggesting that there is no significant variation in the counts of passive and active ellipticals across different environments. The fraction of star-forming galaxies among late-type galaxies is also constant across densities ranging over three orders of magnitude. The fact that this fraction is $\sim$ 1.0 reflects the fact that disk galaxies are mostly (96\%) forming stars and are not in a composite  (AGN plus star-formation) or AGN  stage. (Note that bins where the ratio is $>1.0$ are the consequence of the existence of star-forming early-type galaxies: since we are normalizing by the number of late-type galaxies,  the ratio can be greater than 1.0). The results do not change significantly if we divide the sample into early and late types by requiring the early-type galaxies to have the likelihood of a de-Vaucouleurs profile higher than the likelihood of a disk profile. This suggests that on the scales sampled in this study, the correlation of activity with density is a by-product of the density-morphology correlation (or vice versa).

\begin{figure}[h]
\begin{center}
\includegraphics[scale=0.7,angle=0]{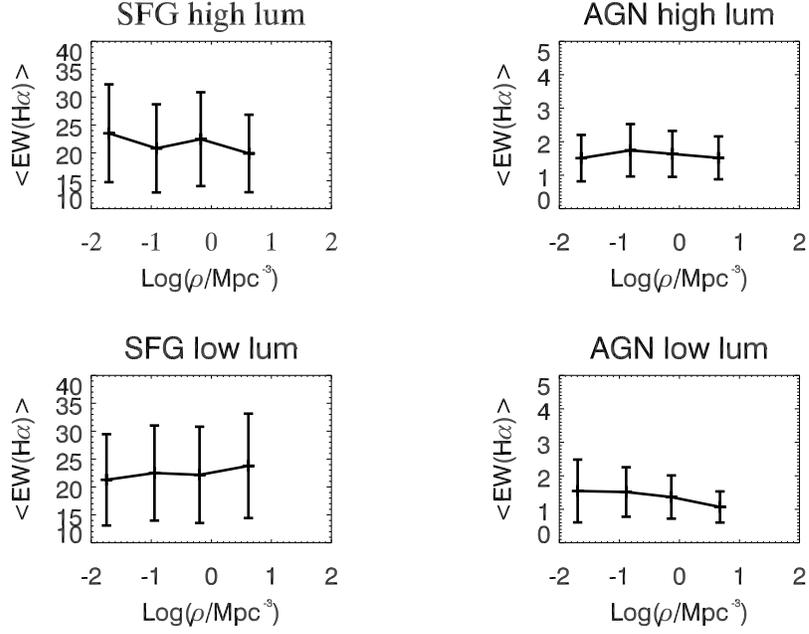}
\end{center}
\caption{\footnotesize \emph{Beyond the density-morphology correlation: the strength of activity in galaxies does not vary significantly with density when the same type of host is chosen. The star-forming sub-samples have been selected from a high-luminosity volume-limited sample ($-21<R<-20$)  and a low-luminosity one ($-19.5<R<-18.5$). Similarly the AGN subsamples have been selected  from a high-luminosity volume limited sample($-22.5<R<-21.5$)  and a low-luminosity volume-limited sample ($-20.5<R<-19.5$). Error bars are the absolute median deviation in each bin. }}
\bigskip
\label{VL_EW_trends}
\end{figure}

 This can be further seen by considering  the intensity of the two types of activity. We  selected  the same type of galaxy in different environments and compared the average intensity of the  associated activity.
Star-formation is typically found in  galaxies with lower absolute $R$ magnitude than the mean absolute $R$ magnitude; we therefore selected two different volume-limited samples at different absolute R magnitudes to maximize the size of the samples. We choose $-21<R<-20$ for star-forming galaxies and $-22.5<R<-21.5$ for AGN, defining these as high-luminosity host samples. We compared these with a low-luminosity sample of host for each class ($-19.5<R<-18.5$ for SFG and $-20.5<R<-19.5$ for AGN).
In Figure \ref{VL_EW_trends} we show the trend for the median equivalent width of H$\alpha$ (EW(H$\alpha$)) with varying density in these different samples spectroscopically classified as AGN or SFG. No significant variation of the  EW(H$\alpha$) is found with  density.
For AGN this suggests that the relative intensity of their emission is not determined by the frequency of their interactions with other galaxies, in agreement with the results of \cite{Nandra2007}.
For SFG the EW(H$\alpha$) measures the ratio of the current to past star-formation activity. Since galaxies in denser environments should have undergone more interactions over time, the invariance of EW(H$\alpha$) with density suggests that the environment plays a relatively small role in shaping their star-formation history, in agreement with other studies  (Carter et al. 2001; Lewis et al. 2002; Hogg et al. 2006).

 Obviously one might argue that the transformation in morphology and activity  may be triggered on much smaller scales by close interactions. However,  to recover the correlation of morphology and activity with the \emph{large-scale} environmental density sampled in this study  requires that, over time, galaxies in denser environments must have undergone more close  interactions than their counterparts in low-density environments or that more galaxies are tidally interacting  in denser environments. If galaxies in denser environments  have undergone more close  interactions than their counterparts in low-density environments, there should be a significant difference in the luminosity-concentration  properties of the populations in different large-scale environments, because the large and the small scale would be coupled. We do not find such a  difference. To assume that  more galaxies in denser environments are in close pair/tidally interacting systems than in low-density environments may explain the increase in the fraction of early- to late-type with increasing large-scale density and the invariance of the luminosity-concentration relation with density, but it seems unlikely to account for  the \emph{smooth} variation of the bulge-to-disk ratio observed in fig. \ref{bulgy_density}. We note, however, that in this study we do not divide the extreme environment of the core of clusters from the bulk of the sample. Whether gas-gas interactions and harassment play a major role in shaping the properties of galaxies in such rare environments cannot be assessed by this study.

\section{Discussion}
\label{discussion_ch3}

Several scenarios have been proposed to explain the variety of
galaxy properties, either by postulating a primordial
difference in early- and late-type galaxies or by assuming an
evolution of late-type galaxies into early types via
mergers, interaction and/or secular evolution. At the
moment, no consensus has been reached on which of these
scenarios is preferable. Certainly  all must play a role in shaping 
the morphology of galaxies and it cannot be excluded, \textit{a priori}, that they are
equally important in creating  the trends observed. 
 
Each of these mechanisms, however, produces a different re-arrangement of gas and angular momentum in galaxies  with its own characteristic  time scale and induced burst of star formation and, potentially, AGN activity. It is therefore likely that understanding the morphological  transformation of galaxies is strongly correlated with understanding a galaxy's star-formation history and the evolution  of its AGN activity.

Coherent information about galaxy  morphology, star-formation,  AGN activity, and supermassive black hole assembly would allow us to  place important constraints on the different mechanisms at play and, possibly, allow us to identify what produces the salient differences in morphology, AGN activity and star-formation history along the Hubble  sequence. The fact that star-formation activity, AGN activity and the growth of bulges and SMBHs  \cite[]{Cowie1996, Cowie2003, Merloni2004} proceed from bigger to smaller systems suggests that galaxies do form a continuum, with more massive galaxies formed at earlier times and evolving faster  than low-mass ones.  In agreement with this idea, the \emph{smooth} scaling relations between morphology, star-formation and AGN activity found in our data seem to confirm a smooth transition in the properties  of galaxies with increasing mass or luminosity.

In this study we have investigated the transformation of similar types of galaxies  in the look back time of the SDSS in order to better understand the mechanisms behind the evolution of galaxies.

The evolution over the past Gyr both in morphology and spectroscopic and radio activity appears more significant for less luminous systems, in agreement with downsizing in the mass assembly and activity observed at higher redshifts \cite[]{Cimatti2006, Cowie2003, Cowie1996}.The observed evolution is therefore likely the tail of the evolution of galaxies and their nuclei observed at higher redshifts.
 Consistent with this notion, we have found that the smooth transition of spectral properties in AGN is accompanied by a different history of star-formation: as the nuclear emission lines become less and less prominent, the last star-formation event dates  to earlier times. This strongly supports a co-evolution of star-formation and AGN activity.
Furthermore we have shown that mid- and low-luminosity hosts of AGN show evidence for a re-arrangement of the stellar component, with the bulge-to-disk ratio increasing within the look-back time of the SDSS, while star-forming systems do not show signatures of significant structural evolution, suggesting that the disk phase (as the elliptical phase) is stable, while the bulgy phase is not and leads to a progressive growth of the bulge. If this phase is short compared to the  red (early-type) and blue (late type) phases, as our study indicates, a shortage of systems with intermediate colors  is to be expected, as suggested by \cite{Balogh2004}. From this point of view, the red sequence/blue sequence dichotomy found in the local population of galaxies \cite[]{Strateva2001} can be interpreted as a simple by-product of the different evolutionary timescales for systems of different mass/luminosity.

The data presented in this study  show that galaxies of different sizes have evolved differently over the past 2 Gyrs. For these changes to be environmentally driven, requires that the effect of the environment should be proportionally weaker on smaller galaxies. This seems unlikely, especially considering the fact that bigger  galaxies tend to inhabit denser region of the large scale structure where the effects of environment (harassment, gas-gas interactions) should be \emph{more} significant.  In agreement with this we have shown that similar galaxies in different environment share similar morphology and activity properties, leading to the conclusion that for  the bulk of galaxies the transformation of both galaxy activity and morphology is mainly a product of the intrinsic properties of the galaxies themselves, not of the environment, which might nonetheless have an important role in the most extreme environments such as the cores of clusters and tidally interacting pairs.

\begin{figure}[h]
\begin{center}
\includegraphics[scale=0.6,angle=0]{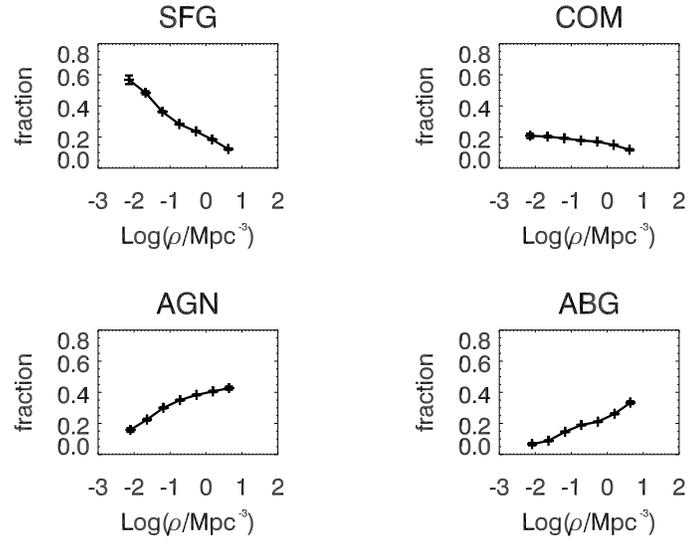}
\end{center}
\caption{\footnotesize \emph{Fractional abundance of the spectroscopic types. Clockwise from the top: star-forming galaxies, composite (star-forming plus AGN) galaxies, passive (absorption-line) galaxies and AGN.}}
\bigskip
\label{fraction_rho_final}
\end{figure}

\begin{figure}
\begin{center}
\includegraphics[scale=0.5,angle=90]{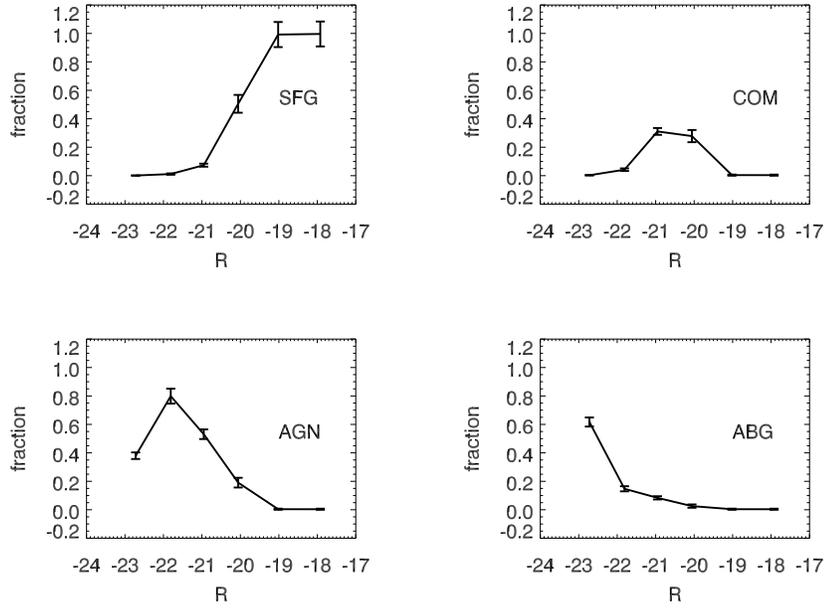}
\end{center}
\caption{\footnotesize \emph{Trend in the fractional abundances of different spectroscopic types with luminosity  }}
\bigskip
\label{HS_fractions}
\end{figure}

From this point of view the most likely factor influencing the evolution of galaxies is the depth of the potential well: in this case more massive galaxies are expected to be more evolved systems.  As shown in  figure \ref{HS_fractions}, star-forming galaxies peak    at absolute magnitudes of $R \sim$-18, composite systems at  $R \sim$ -20.5, AGN at $R \sim$ -21.5 and passive systems at $R \sim$ -23 in agreement with a scenario in which galaxies of different mass/luminosity are at different stages in their evolution from star-forming to passive systems, passing through an intermediate AGN phase. Since more massive/luminous systems reside in denser environments, the trends in figure \ref{fraction_rho_final} would be a  consequence of the different rate of evolution for systems of different mass.

The simplest way to interpret our data is therefore  to posit
that galaxies of different sizes underwent evolution at  different rates, with smaller systems still in the earlier stages. 

We note that the infall of baryonic matter in a potential well scales as $t\propto \rho^{-0.5}$. Therefore bigger primordial over-densities should have faster infall of matter, while small over-densities should have a much slower infall. This may be the key in understanding galaxy formation and evolution. 

A fast infall/collapse  of material in bigger over-densities would more likely resemble a monolithic collapse. These over-densities  would form big bulges and big ellipticals (and supermassive black holes within). A slow accretion phase typical of small over-densities  would most likely find a thin disk  equilibrium configuration, forming smaller bulge-less spiral galaxies over longer timescales. Galaxies with intermediate potential wells would have both.

While disks and spheroids are mostly  stable configurations, the mix of the two may not be. As the gas continues to fall into the potential well, the disk-to-bulge ratio continues to vary, as we have shown in this study. As gas is transferred from the outer regions into the inner regions of the potential well, AGN activity is powered. As the  reservoir of cold gas  gets depleted, the AGN strength decreases as we have shown. Strong AGN in the local universe would  be associated with bulgy systems with plenty of gas that at earlier times where mostly star-forming systems, as observed.

\begin{figure}
\begin{center}
\includegraphics[scale=0.5,angle=90]{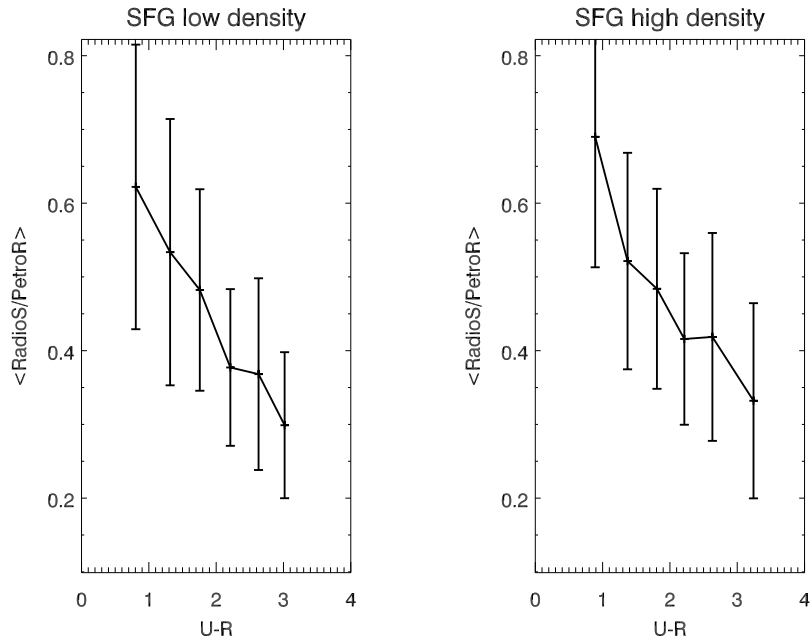}
\end{center}
\caption{\footnotesize \emph{The median concentration of the radio emission in star-forming galaxies in high ($\rho>1$ galaxy/Mpc$^{3}$) and low ($\rho<1$ galaxy/Mpc$^{3}$) density environments. The emission becomes more centrally concentrated in redder galaxies, and the  concentration does not depend on the environment. Radio sizes have been determined as the HWHM of a two-dimensional gaussian fit to the FIRST radio maps. }}
\bigskip
\label{color_size}
\end{figure}

Furthermore, if the gaseous reservoir of
galaxies is progressively transferred into the inner
regions, one would expect that star-formation would
become more and more centrally concentrated as a
galaxy ages.  We have explored this by
considering the ratio of the radio to optical sizes of our galaxies and
have found that this ratio decreases for redder systems,
consistent with the notion that star-formation becomes more and more
centrally concentrated for more evolved types  (Figure \ref{color_size}), in agreement with the early work of Hummel (1981). Such a  redistribution of the gas is  in agreement with the findings of \cite{Cayatte1994}, who showed a decreasing ratio of HI to optical radii along the Hubble sequence.  Recently, evidence of  gas infall in the inner regions of late-type galaxies produced by secular evolution has been demonstrated by \cite{Regan2006}. 

This type of scenario seems to favor  a transformation of the bulk of galaxies given by the combined effect of the  progressive  infalling of matter from the outer regions of galaxies \cite[]{Fraternali2001}  and its re-arrangement  by means of disk instabilities such as bars and density waves \cite[]{Combes2000, Zhang1996, Sellwood2002, Kormendy2004}.
 Since the properties of the gas infall mostly depends on the potential well of the galaxy, the properties of galaxies must  vary smoothly with increasing luminosity and objects of different luminosity should have different signatures of activity, as we find  in the data. While this type of scenario would naturally predict downsizing and account for several observational properties, it stands in contrast to CDM models, since it would require the over-densities associated with massive ellipticals to be formed at an early time, and not hierarchically. 
A merger-driven transformation of galaxies , on the contrary, seems unlikely. First, the evolution appears to be anti-hierarchical, with smaller systems evolving more at later times. Secondly, mergers seem unlikely  to produce  the  smooth transformation of both star-formation \emph{and} AGN activity in hosts of different masses described in this paper. Furthermore, since such a mechanism depends on the density of the environment and the relative velocity of the encounters, it is unlikely that it can work equally well in all environments as required by Figure  \ref{morpho_lum}. Also, it is not plausible that mergers would affect galaxies of different luminosities in different ways as we find, transforming more low-luminosity systems in the past Gyr than high-luminosity ones. The existence of AGN activity in non-interacting systems \cite[]{Pierce2007} raises further questions for this mechanism.

From this point of view, the bulk of the density-morphology relation might simply be a size-density relation, with bigger disk galaxies preferentially formed  in denser environments at earlier epochs  and evolving more at earlier times than small size galaxies.  Late-type galaxies in denser environments are then expected to have more concentrated gas than similar galaxies in less dense environments. Interestingly, \cite{Cayatte1994} compared late-type galaxies in the Virgo cluster with galaxies in the field and found that HI  is more concentrated in cluster galaxies (although they interpreted  this as a possible effect of the cluster environment on the Virgo galaxies, such as stripping).

The size-density correlation would lead to the correlation of star-formation and AGN activity with density shown in figure \ref{fraction_rho_final}: smaller galaxies which inhabit less dense environments have longer transformation  times and therefore, at the present epoch, are still forming stars, while bigger galaxies have evolved into spheroids and their gaseous reservoir is only used to feed the SMBH at their centers. When the trends are normalized respectively to late- and early-type galaxies, the correlation of activity with environment disappears: there is no substantial difference in the ratio of star-forming disk galaxies or AGN early-type galaxies  across the range in densities across the large scale structure (Figure \ref{sfg_late_agn_early}). These results are in agreement with earlier work by \cite{Schmitt2001}.  From this point of view, understanding the density-morphology and the density-activity correlation would mean understanding why  bigger galaxy-scale over-densities were preferentially born (or have aggregated) within  denser large-scale environments.

\section{Conclusion}
\label{conclusion_ch3}

In this study based on the SDSS and FIRST Surveys, we have investigated the evolution of galaxy properties withing the SDSS look-back time ($\sim$ 2 Gyr) for a sample of  $\sim$ 150000 galaxies in the local universe.
We have shown a transformation of the activity and morphology of galaxies over the look-back time of the SDSS and have further demonstrated that this transformation is more significant for lower luminosity systems, in agreement with the notion of downsizing  in star formation, AGN activity and bulge formation over cosmic time.  The intensity of the H$\alpha$ and radio emission has significantly dropped over the last 1 Gyr in emission-line AGN, a fact in agreement with the progressive fading of active nuclei from high to low redshift traced by other surveys. The hosts of these nuclei have substantially transformed towards more concentrated stellar distributions typical of more bulgy systems.
The fact that systems of different luminosity evolve differently over this look-back time supports the claim that, for the bulk of galaxies, transformation is unlikely the product of environment-driven mechanisms, but rather depends on the mass/size of the host itself.

This is further supported by the direct study of the galaxy environment presented in this paper. 
We have shown that galaxies show increasing bulge-to-disk ratio with increasing density and that this increase is accompanied by a transition in the type of activity from star-formation-dominated to composite to AGN-dominated and eventually to passive systems. This transition accompanies the well-know density-morphology relation, into regions well beyond the cluster cores where it was originally discovered (Dressler 1980).  We argue that these two trends with densities  are unlikely driven by  environment-driven mechanisms,  but rather depend on the mass/size of the host itself and different evolution timescales for galaxies associated to different potential wells.

We note that if over-densities of all sizes were formed at early epochs instead of being hierarchically formed over time, the infall time of gas would be much faster for the largest over-densities and correspondingly slower for the smaller ones.
A fast infall/collapse  of material in bigger over-densities would more likely resemble a monolithic collapse. These over-densities  would form big bulges and big ellipticals (and supermassive black holes within). A slow accretion phase typical of small over-densities  would most likely find a thin disk  equilibrium configuration, forming smaller bulge-less spiral galaxies over longer timescales. Galaxies with intermediate potential wells would have both and be a less stable configuration. 
In agreement with this we have shown that big bulge-dominated galaxies and star-forming disks are stable configurations, while galaxies with intermediate bulges  hosting AGN tend to evolve towards  bulgier systems. This rapid transformation of systems of intermediate luminosity into earlier type systems with more prominent bulges demonstrated here, suggests that the intermediate phase may be  short-lived compared to the star-forming disk phase and the red elliptical phase, thus explaining the dichotomy between red- and blue-sequence galaxies. 

The progressive infall of gas from the outer regions to the inner regions of a galaxy potential well can be traced using the size of radio emission as an indicator of the star-formation region within galaxies. We have shown that  a progressive concentration of star-formation activity towards the inner regions of redder star-forming galaxies is found, in agreement with early work of Hummels (1981), suggesting a picture in which gas is progressively transported by instabilities towards the inner regions. As the gas is  funneled towards the center, star-formation is triggered, followed by nuclear activity. As this cold gas reservoir is exhausted the AGN activity is progressively brought to a halt, explaining the anti-correlation that we find between line-strength of active nuclei and galaxy color (and other indicators of star-formation activity), which is an extension of the results of \cite{Kauffmann2003}.

 From this point of view the density-morphology relation and the density-activity relation discussed in this study is the by-product of different formation and evolution timescales for galaxies associated with different primordial over-densities, and the fact that  the biggest galaxy-scale over-densities were formed within the denser large-scale structure over-densities.

\section{Acknowledgments}

This work greatly benefited from discussion with Jacqueline van Gorkom, Ed Spiegel, David Schiminovich, David Hogg and Margaret Geller.

PMR was supported by the National Science Foundation under grant AST-06-07643.

Funding for Sloan Digital Sky Survey project has been provided  by the Alfred P. Sloan Foundation, the Participating Institutions, the National Aeronautics and Space Administration, the National Science Foundation, the U.S. Department of Energy, the Japanese Monbukagakusho, and the Max Planck Society. The SDSS website is http://www.sdss.org .

The SDSS is managed by the Astrophysical Research Consortium (ARC) for the Participating Institutions (The University of Chicago, Fermilab, the Institute for Advanced Study, the Japan Participation Group, The Johns Hopkins University, the Korean Scientist Group, Los Alamos National Laboratory, the Max-Planck-Institute for Astronomy, the Max-Planck-Institute for Astrophysics, New Mexico State University, University of Pittsburgh, University of Portsmouth, Princeton University, the United States Naval Observatory, and the University of Washington).

\clearpage 

\clearpage

\clearpage

\clearpage

\clearpage

\clearpage

\clearpage

\clearpage

\clearpage

\clearpage

\clearpage

\clearpage

\clearpage

\clearpage

\clearpage

\clearpage

\clearpage

\clearpage

\clearpage

\clearpage

\clearpage

\clearpage

\clearpage

\clearpage

\clearpage

\clearpage

\clearpage

\clearpage


\begin{thebibliography}

\bibitem[Balogh et al.(2004)]{Balogh2004} Balogh, M.~L., Baldry, 
I.~K., Nichol, R., Miller, C., Bower, R., \& Glazebrook, K.\ 2004, \apjl, 
615, L101
\bibitem[Barton, Geller, \& Kenyon(2000)]{Barton2000} Barton, 
E.~J., Geller, M.~J., \& Kenyon, S.~J.\ 2000, \apj, 530, 660 
\bibitem[Becker, White, \& Helfand(1995)]{Becker95} Becker, 
R.~H., White, R.~L., \& Helfand, D.~J.\ 1995, \apj, 450, 559 
\bibitem[Best(2004)]{Best2004} Best, P.~N.\ 2004, \mnras, 351, 
70 
\bibitem[Best et al.(2005)]{Best2005} Best, P.~N., Kauffmann, 
G., Heckman, T.~M., \& Ivezi{\'c}, {\v Z}.\ 2005, \mnras, 362, 9 
\bibitem[Brinchmann et al.(2004)]{Brinchmann2004} Brinchmann, J., 
Charlot, S., White, S.~D.~M., Tremonti, C., Kauffmann, G., Heckman, T., \& 
Brinkmann, J.\ 2004, \mnras, 351, 1151 
\bibitem[Bruzual A.(1983)]{Bruzual1983} Bruzual A., G.\ 1983, \apj, 
273, 105 
\bibitem[Butcher \& Oemler(1978)]{Butcher1978} Butcher, H., \& 
Oemler, A., Jr.\ 1978, \apj, 226, 559 
 \bibitem[Carter et al.(2001)]{Carter01} Carter, B.J., Fabricant, D.G., Geller, M.J., \& Kurtz, M.J.,2001,\apj,559,606
\bibitem[Cassata et al.(2007)]{Cassata2007} Cassata, P., et al.\ 
2007, \apjs, 172, 270 
\bibitem[Cayatte et al.(1994)]{Cayatte1994} Cayatte, V., Kotanyi, 
C., Balkowski, C., \& van Gorkom, J.~H.\ 1994, \aj, 107, 1003 
\bibitem[Christlein(2000)]{Christlein2000} Christlein, D.\ 2000, \apj, 
545, 145 
\bibitem[Christlein 
\& Zabludoff(2005)]{Christlein2005} Christlein, D., \& Zabludoff, A.~I.\ 2005, \apj, 621, 201 
\bibitem[Cimatti et al.(2006)]{Cimatti2006} Cimatti, A., Daddi, E., 
\& Renzini, A.\ 2006, \aap, 453, L29 
\bibitem[Combes(2000)]{Combes2000} Combes, F.\ 2000, Dynamics of 
Galaxies: from the Early Universe to the Present, 197, 15 
\bibitem[Combes(2006)]{Combes2006} Combes, F.\ 2006, Astrophysics 
Update 2, 159 
\bibitem[Condon(1992)]{Condon92} Condon, J.~J.\ 1992, \araa, 30, 
575 
\bibitem[Condon et al.(1993)]{Condon1993} Condon, J.~J., Cotton, 
W.~D., Greisen, E.~W., Perley, R.~A., Yin, Q.~F., \& Broderick, J.~J.\ 
1993, Bulletin of the American Astronomical Society, 25, 1389 
\bibitem[Cooper et al.(2008)]{Cooper2008} Cooper, M.~C., et al.\ 
2008, \mnras, 383, 1058 
\bibitem[Cowie et al.(1996)]{Cowie1996} Cowie, L.~L., Songaila, 
A., Hu, E.~M., \& Cohen, J.~G.\ 1996, \aj, 112, 839 
\bibitem[Cowie et al.(2003)]{Cowie2003} Cowie, L.~L., Barger, 
A.~J., Bautz, M.~W., Brandt, W.~N., \& Garmire, G.~P.\ 2003, \apjl, 584, L57 
\bibitem[Dahari(1984)]{Dahari1984} Dahari, O.\ 1984, \aj, 89, 966 
\bibitem[Dressler(1980)]{Dressler1980} Dressler, A.\ 1980, \apj, 
236, 351 
\bibitem[Dunlop et al.(2003)]{Dunlop2003} Dunlop, J.~S., McLure, 
R.~J., Kukula, M.~J., Baum, S.~A., O'Dea, C.~P., \& Hughes, D.~H.\ 2003, 
\mnras, 340, 1095 
\bibitem[Ferrarese \& Merritt(2000)]{Ferrarese2000} Ferrarese, L., \& 
Merritt, D.\ 2000, \apjl, 539, L9 
\bibitem[Fraternali et al.(2001)]{Fraternali2001} Fraternali, F., 
Oosterloo, T., Sancisi, R., \& van Moorsel, G.\ 2001, \apjl, 562, L47 
\bibitem[Geller et al.(2006)]{Geller2006} Geller, M.~J., Kenyon, 
S.~J., Barton, E.~J., Jarrett, T.~H., \& Kewley, L.~J.\ 2006, \aj, 132, 2243 
\bibitem[Glikman et al.(2004)]{Glikman2004} Glikman, E., Helfand, D.~J., Becker, R.~H., \& White, R.~L.\ 2004, ASP Conference Series 311: AGN Physics with the Sloan Digital Sky Survey 311, 351 
\bibitem[G{\' o}mez et al.(2003)]{Gomez2003} G{\' o}mez, P.~L., 
et al.\ 2003, \apj, 584, 210 
\bibitem[Grimm et al.(2003)]{Grimm2003} Grimm, H.-J., Gilfanov, 
M., \& Sunyaev, R.\ 2003, \mnras, 339, 793 
\bibitem[Grogin et al.(2005)]{Grogin2005} Grogin, N.~A., et al.\ 
2005, \apjl, 627, L97 
\bibitem[Helou, Soifer, \& Rowan-Robinson(1985)]{Helou1985} 
Helou, G., Soifer, B.~T., \& Rowan-Robinson, M.\ 1985, \apjl, 298, L7 
\bibitem[Hogg et al.(2004)]{Hogg2004} Hogg, D.~W., et al.\ 2004, 
\apjl, 601, L29 
\bibitem[Hogg et al.(2006)]{Hogg2006} Hogg, D.~W., Masjedi, M., 
Berlind, A.~A., Blanton, M.~R., Quintero, A.~D., 
\& Brinkmann, J.\ 2006, \apj, 650, 763 
\bibitem[Hopkins et al.(2006)]{Hopkins2006} Hopkins, P.~F., 
Hernquist, L., Cox, T.~J., Di Matteo, T., Robertson, B., 
\& Springel, V.\ 2006, \apjs, 163, 1 
\bibitem[Hubble(1936)]{Hubble1936} Hubble, E.~P.\ 1936, Yale 
University Press,  
\bibitem[Kauffmann et al.(2003)]{Kauffmann2003} Kauffmann, G., et 
al.\ 2003, \mnras, 346, 1055 
\bibitem[Kauffmann et al.(2003)]{Kauffmann2003b} Kauffmann, G., et 
al.\ 2003, \mnras, 341, 54 
\bibitem[Kauffmann et al.(2004)]{Kauffmann2004} Kauffmann, G., White, 
S.~D.~M., Heckman, T.~M., M{\' e}nard, B., Brinchmann, J., Charlot, S., 
Tremonti, C., \& Brinkmann, J.\ 2004, \mnras, 353, 713 
\bibitem[Kriek et al.(2007)]{Kriek2007} Kriek, M., et al.\ 2007, 
\apj, 669, 776 559 
\bibitem[Kormendy \& Kennicutt(2004)]{Kormendy2004} Kormendy, J., \& 
Kennicutt, R.~C., Jr.\ 2004, \araa, 42, 603 
\bibitem[Lewis et al.(2002)]{Lewis2002} Lewis, I., et al.\ 2002, 
\mnras, 334, 673 
\bibitem[Longair(1966)]{Longair1966} Longair, M.~S.\ 1966, \mnras, 
133, 421 
\bibitem[Lynden-Bell \& Kalnajs(1972)]{Lynden1972} Lynden-Bell, 
D., \& Kalnajs, A.~J.\ 1972, \mnras, 157, 1 
\bibitem[Madau et al.(1998)]{Madau1998} Madau, P., Pozzetti, L., 
\& Dickinson, M.\ 1998, \apj, 498, 106 
\bibitem[Martini et al.(2007)]{Martini2007} Martini, P., Mulchaey, 
J.~S., \& Kelson, D.~D.\ 2007, \apj, 664, 761 
\bibitem[Mateus \& Sodr{\' e}(2004)]{Mateus2004} Mateus, A.~\& 
Sodr{\' e}, L.\ 2004, \mnras, 349, 1251 
\bibitem[Meier(2001)]{Meier2001} Meier, D.~L.\ 2001, 20th Texas 
Symposium on relativistic astrophysics, 586, 420 
\bibitem[Merloni et al.(2004)]{Merloni2004} Merloni, A., Rudnick, 
G., \& Di Matteo, T.\ 2004, \mnras, 354, L37 
\bibitem[Miller et al.(2003)]{Miller2003} Miller, C.~J., Nichol, 
R.~C., G{\' o}mez, P.~L., Hopkins, A.~M., \& Bernardi, M.\ 2003, \apj, 597, 
142 

\bibitem[Moshir et al.(1990)]{Moshir1990} Moshir, M., et al.\ 
1990, \baas, 22, 1325
 
\bibitem[Nandra et al.(2007)]{Nandra2007} Nandra, K., et al.\ 
2007, \apjl, 660, L11
\bibitem[Park et al.(2007)]{Park2007} Park, C., Choi, Y.-Y., 
Vogeley, M.~S., Gott, J.~R.~I., \& Blanton, M.~R.\ 2007, \apj, 658, 898 
 \bibitem[Pierce et al.(2007)]{Pierce2007} Pierce, C.~M., et al.\ 
2007, \apjl, 660, L19 
\bibitem[Poggianti et al.(2001)]{Poggianti2001} Poggianti, B.~M., et 
al.\ 2001, \apj, 563, 118 
\bibitem[Postman \& Geller(1984)]{Postman1984} Postman, M., \& 
Geller, M.~J.\ 1984, \apj, 281, 95 
\bibitem[Ranalli et al.(2003)]{Ranalli2003} Ranalli, P., Comastri, 
A., \& Setti, G.\ 2003, \aap, 399, 39
\bibitem[Regan et al.(2006)]{Regan2006} Regan, M.~W., et al.\ 
2006, \apj, 652, 1112 
\bibitem[Reviglio(2003)]{Reviglio2003} Reviglio P., ``Multiwavelength Analysis of Star-Forming Galaxies and Active Galactic Nuclei in a Complete Redshift Survey'', Tesi di Dottorato di Ricerca,  Universit\`a di Torino, XVI Ciclo, 2001-2003 
\bibitem[Reviglio \& Helfand(2006)]{Reviglio2006} Reviglio, P., \& 
Helfand, D.~J.\ 2006, \apj, 650, 717
\bibitem[Reviglio \& Helfand(2009)]{Reviglio2009} Reviglio, P., \& 
Helfand, D.~J.\ 2007, in publication, Paper I
\bibitem[Reviglio(2008)]{Reviglio2008} Reviglio, P., ``Active Galaxies and their Evolution as observed in the FIRST and Sloan Digital Sky Surveys'', Ph.D. Thesis, Columbia University, New York, U.S.A., 2008
\bibitem[Rieke \& Lebofsky(1985)]{Rieke1985} Rieke, G.~H., \& Lebofsky, M.~J.\ 1985, \apj, 288, 618 
\bibitem[Schmitt(2001)]{Schmitt2001} Schmitt, H.~R.\ 2001, \aj, 
122, 2243 
\bibitem[Sellwood 
\& Binney(2002)]{Sellwood2002} Sellwood, J.~A., \& Binney, J.~J.\ 2002, \mnras, 336, 785 
\bibitem[Strateva et al.(2001)]{Strateva2001} Strateva, I., et al.\ 
2001, \aj, 122, 1861 
\bibitem[Ulvestad \& Ho(2001)]{Ulvestad2001} Ulvestad, J.~S., \& Ho, 
L.~C.\ 2001, \apj, 558, 561 
 \bibitem[Veilleux \& Osterbrock(1987)]{Veilleux1987} Veilleux, S.~\& 
Osterbrock, D.~E.\ 1987, \apjs, 63, 295 
\bibitem[Voges et al.(1999)]{Voges1999} Voges, W., et al.\ 1999, 
\aap, 349, 389 
\bibitem[Voges et al.(2000)]{Voges2000} Voges, W., et al.\ 2000, 
VizieR Online Data Catalog, 9029, 0 
\bibitem[White et al.(2006)]{White2006} White, R.~L., Helfand, 
D.~J., Becker, R.~H., Glikman, E., \& deVries, W.\ 2006, ArXiv Astrophysics 
e-prints, arXiv:astro-ph/0607335 
\bibitem[White, Giommi, \& Angelini(2000)]{White2000} White, 
N.~E., Giommi, P., \& Angelini, L.\ 2000, VizieR Online Data Catalog, 9031, 
0 
\bibitem[White et al.(2007)]{White2007} White, R.~L., Helfand, 
D.~J., Becker, R.~H., Glikman, E., \& de Vries, W.\ 2007, \apj, 654, 99 
\bibitem[Worthey 
\& Ottaviani(1997)]{Worthey1997} Worthey, G., \& Ottaviani, D.~L.\ 1997, \apjs, 111, 377 
\bibitem[York et al.(2000)]{York2000} York, D.~G., et al.\ 2000, 
\aj, 120, 1579 

\bibitem[Zhang(1996)]{Zhang1996} Zhang, X.\ 1996, \apj, 457, 125 
\bibitem[Zirbel \& Baum(1995)]{Zirbel1995} Zirbel, E.~L., \& Baum, S.~A.\ 1995, \apj, 448, 521 

\end{thebibliography}
\end{document}